\documentclass[aps,prb,twocolumn,superscriptaddress,floatfix]{revtex4-1}

\usepackage{graphicx}
\usepackage{eqnarray,amsmath}
\usepackage{amsthm}
\usepackage{dsfont}
\usepackage{amssymb,color}
\usepackage[caption=false]{subfig}
\makeatletter
\newcommand*\dashline{\rotatebox[origin=c]{90}{$\dabar@\dabar@\dabar@$}}
\makeatother

\pdfoutput=1

\begin{document}

\title[]{Quasielectrons in lattice Moore-Read models}

\author{Sourav Manna}
\affiliation{Max-Planck-Institut f\"ur Physik komplexer Systeme, D-01187 Dresden, Germany}

\author{Julia Wildeboer}
\affiliation{Max-Planck-Institut f\"ur Physik komplexer Systeme, D-01187 Dresden, Germany}
\affiliation{Department of Physics \& Astronomy, University of Kentucky, Lexington, Kentucky 40506-0055, USA}

\author{Anne E. B. Nielsen}
\affiliation{Max-Planck-Institut f\"ur Physik komplexer Systeme, D-01187 Dresden, Germany}
\affiliation{Department of Physics and Astronomy, Aarhus University, DK-8000 Aarhus C, Denmark}

\begin{abstract}
Moore-Read states can be expressed as conformal blocks of the underlying rational conformal field theory, which provides a well explored description for the insertion of quasiholes. It is known, however, that quasielectrons are more difficult to describe in continuous systems, since the natural guess for how to construct them leads to a singularity. In this work, we show that the singularity problem does not arise for lattice Moore-Read states. This allows us to construct Moore-Read Pfaffian states on lattices for filling fraction $5/2$ with both quasiholes and quasielectrons in a simple way. We investigate the density profile, charge, size and braiding properties of the anyons by means of Monte Carlo simulations. Further we derive an exact few-body parent Hamiltonian for the states. Finally, we compare our results to the density profile, charge and shape of anyons in the Kapit-Mueller model by means of exact diagonalization.
\end{abstract}

\maketitle

\section{Introduction}

Quasiparticles in fractional quantum Hall systems attract a great deal of interest since they carry fractional charge and obey fractional quantum statistics. They are neither bosonic nor fermionic in nature but are {\it anyonic}. Lattice versions of the fractional quantum Hall models are interesting in the field of ultracold atoms in optical lattices\cite{Others49}. The lattice models reveal exotic features which are absent in the continuum systems \cite{Others51}. This might even show the way to realize fractional quantum Hall physics at room temperature\cite{Others48}. Most of the fractional quantum Hall states exhibit Abelian anyons, i.e.\ under quasiparticle exchange the anyonic state acquires only a phase factor $e^{i\phi} \neq \pm 1$. A more remarkable scenario, however, happens when the ground state with fixed quasiparticle positions is degenerate and an exchange of the $i$th and $j$th anyons leads to a unitary transformation $\mathcal{U}_{ij}$ known as the monodromy matrix. If $\mathcal{U}_{ij}$ corresponding to different exchanges do not commute then the statistics are non-Abelian. In the field of quantum information the use of non-Abelian braiding statistics to make qubits in topological quantum computation \cite{C.Nayak9} is attracting much attention.

Quasiparticles are of two types with equal importance, namely {\it quasiholes} and {\it quasielectrons}. Fractional quantum Hall states containing quasiholes are well explored in both Abelian and non-Abelian states,\cite{C.Nayak8, BAB14, DA1, Others52} but the theory of quasielectrons in fractional quantum Hall states turns out to be a more complex problem. The reason behind that is that a singularity appears in the state if one tries to construct the quasielectrons as the inverse of the quasiholes in the continuum. This happens because by inserting flux tubes with positive flux one can obtain a quasihole state, but similarly flux tubes with negative flux do not give rise to a quasielectron state, rather it creates a singularity. This has initiated a lot of work to develop methods and to write proper states for quasielectrons in the continuum \cite{NR3,NR5, Others41, BAB5,FDMH1, Others42, Others43, JKJ7}. These states are, however, rather complicated, and this makes it difficult to investigate quasielectron properties. 

Recently, it was discovered that the singularity does not appear for fractional quantum Hall states defined on lattices.\cite{Anne2} So far this observation has been used to investigate Laughlin quasielectrons in lattice systems in great detail. In the present paper, we investigate the important and more challenging case of non-Abelian quasielectrons in Moore-Read states defined on lattices. 

Moore and Read introduced the Moore-Read states as conformal blocks of the underlying conformal field theory (CFT), and they also showed how one can introduce quasiholes in the states.\cite{Moore-Read1} Here we consider the corresponding states on lattices, and we show that quasielectrons can be introduced in a way parallel to the way quasiholes are introduced. In particular, this means that the wavefunctions containing quasielectrons are not more complicated than wavefunctions containing only quasiholes. This allows us to investigate the properties of the quasielectrons in great detail with Monte Carlo simulations. We present detailed results for the density profile, charge, size and braiding properties of the anyons. In addition, we derive exact parent Hamiltonians for these states.

The Kapit-Mueller model provides a relatively simple Hamiltonian \cite{Others47} for hardcore bosons on a lattice, whose ground state space is in the same topological phase as the Moore-Read state. We consider a lattice with 24 sites and introduce quasielectrons by adding a potential. We find that the size of the quasielectrons is roughly the same as for the corresponding analytical Moore-Read state. This shows that the analytical Moore-Read states are helpful to gain insight into the physics.

The structure of the paper is as follows: In Sec.\ \ref{MR}, we investigate the density profile, charge, shape and braiding statistics of the non-Abelian Ising quasielectrons in the lattice Moore-Read states. We also derive exact parent Hamiltonians for the states. In Sec.\ \ref{KM_Model}, we investigate quasielectrons in the Kapit-Mueller model and show that the size of the quasielectrons is similar to that of the quasielectrons in the analytical Moore-Read state. We conclude the paper in Sec.\ \ref{concl}.

\section{Non-Abelian quasielectrons in the Moore-Read states}\label{MR}

We exploit CFT in lattice systems to describe Moore-Read states containing Ising anyons and use Monte Carlo simulations to investigate the density profile, charge, 
shape and braiding properties of the anyons in detail. Later we will provide parent Hamiltonians for these states. 
We commence by constructing the states from conformal field correlators.

\subsection{Lattice Moore-Read states from conformal field correlators}
For arbitrary lattices in a two-dimensional complex plane with $N$ lattice sites at positions $\{z_j\}, \ j \in \{1,..,N\}$ and $Q$ quasiholes at positions $\{w_k\}, \ k \in \{1,..,Q\}$, 
we associate the two vertex operators \cite{Anne8}
\begin{equation}\label{lattice_operator}
\mathcal{V}_{n_j}(z_j) = e^{i\pi(j-1)\eta n_j}\psi(z_j)^{{\Delta_{n_j}}}:e^{i(qn_j-\eta)\phi(z_j)/\sqrt{q}}:  
\end{equation}
and 
\begin{equation}\label{anyon_operator}
W_{p_k}(w_k) = \sigma(w_k) :e^{i p_k \phi(w_k)/\sqrt{q}}: 
\end{equation}
to each lattice site and quasihole position, respectively. Here, $\phi(z_j)$ is the chiral field of a free massless boson, the inverse filling fraction is $q \in \mathbb{N}$, 
$: \cdots :$ is the normal ordering, and  
$n_j \in \{0,1\}$ is the occupation at site $j$ depicting hardcore bosons (fermions) for $q$ odd (even). We have $\Delta_{n_j} = 1$ iff $n_j = 1$ and otherwise $\Delta_{n_j} = 0$. We define the parameter $\eta = a/2\pi$ with $a$ as the average area per 
lattice site, $\psi(z_j)$ is the majorana field at the occupied sites only ($\Delta_{n_j} = 1$), $\sigma(w_k)$ is the holomorphic spin operator of the chiral Ising CFT and $p_k/q$, 
with $p_k$ positive, is the charge of the quasihole at position $w_k$. 
 
\begin{figure*}
        \includegraphics[width=0.245\textwidth]{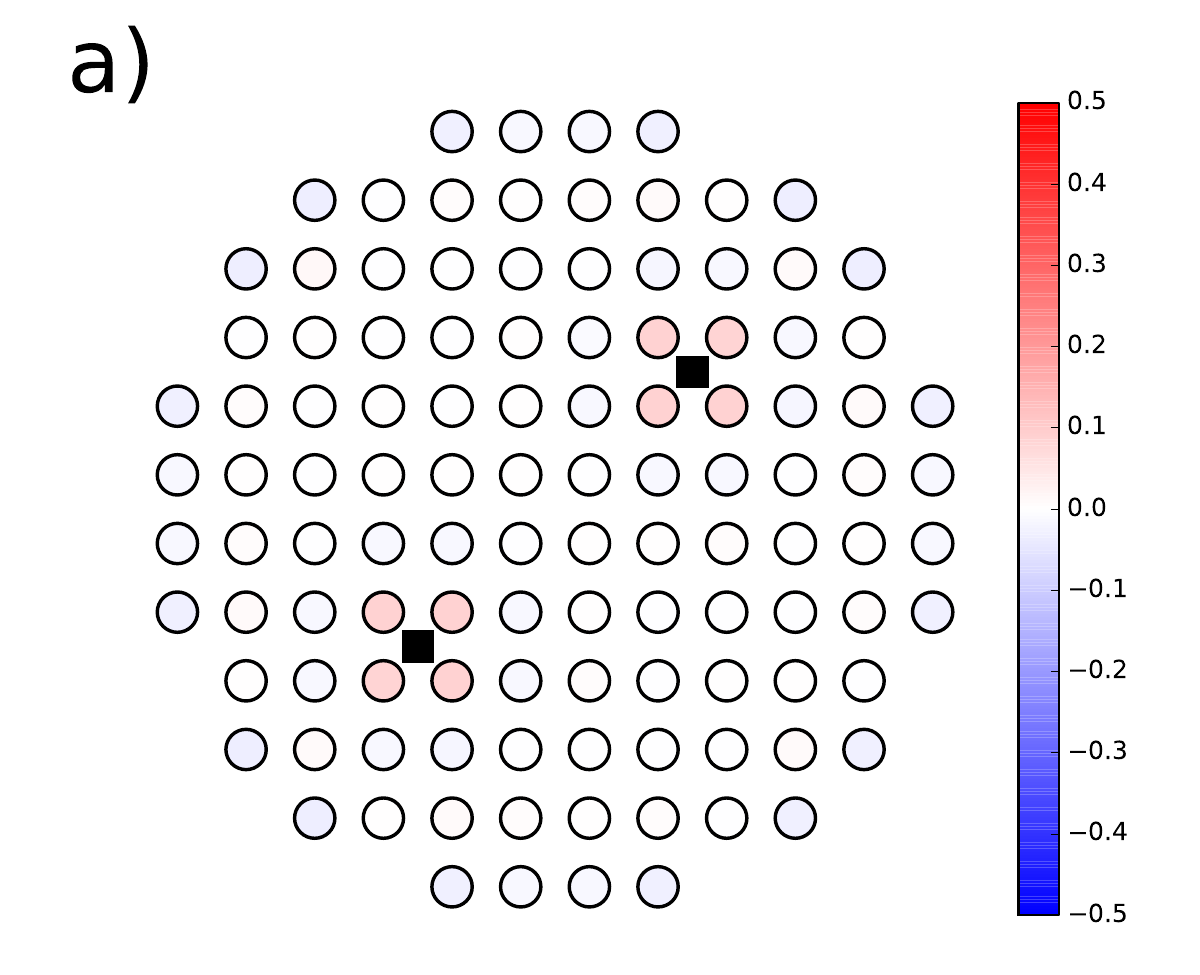}
        \includegraphics[width=0.245\textwidth]{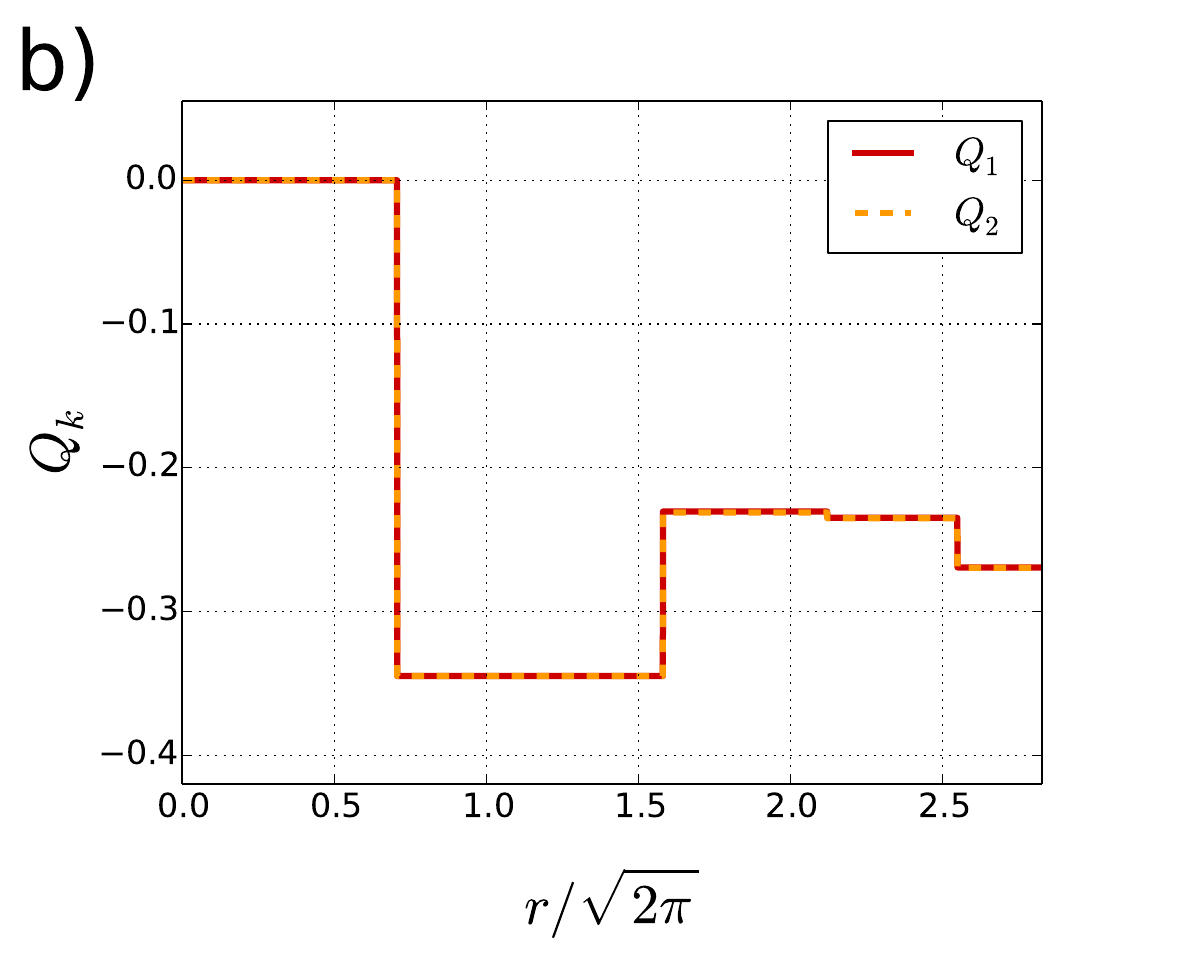}
        \includegraphics[width=0.245\textwidth]{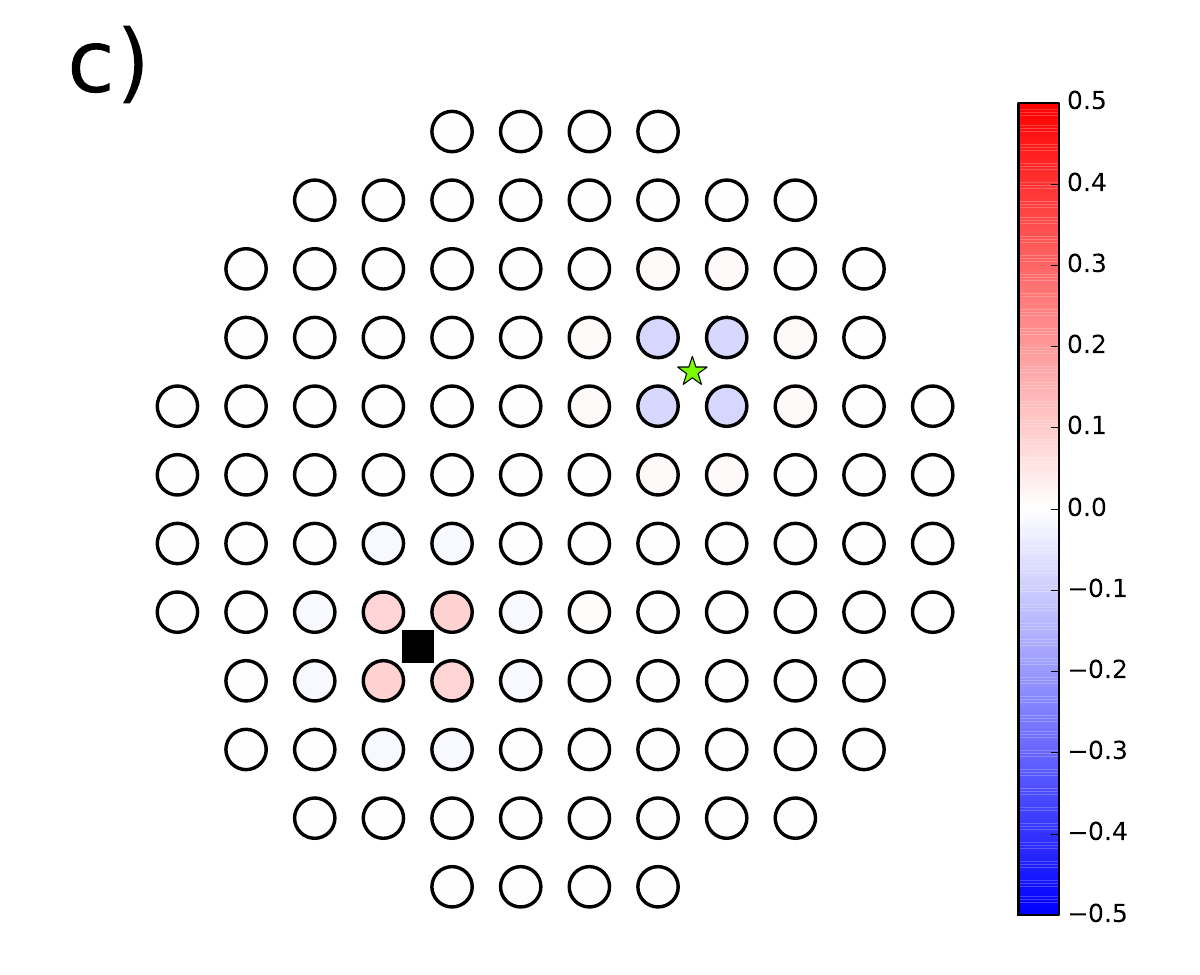}
        \includegraphics[width=0.245\textwidth]{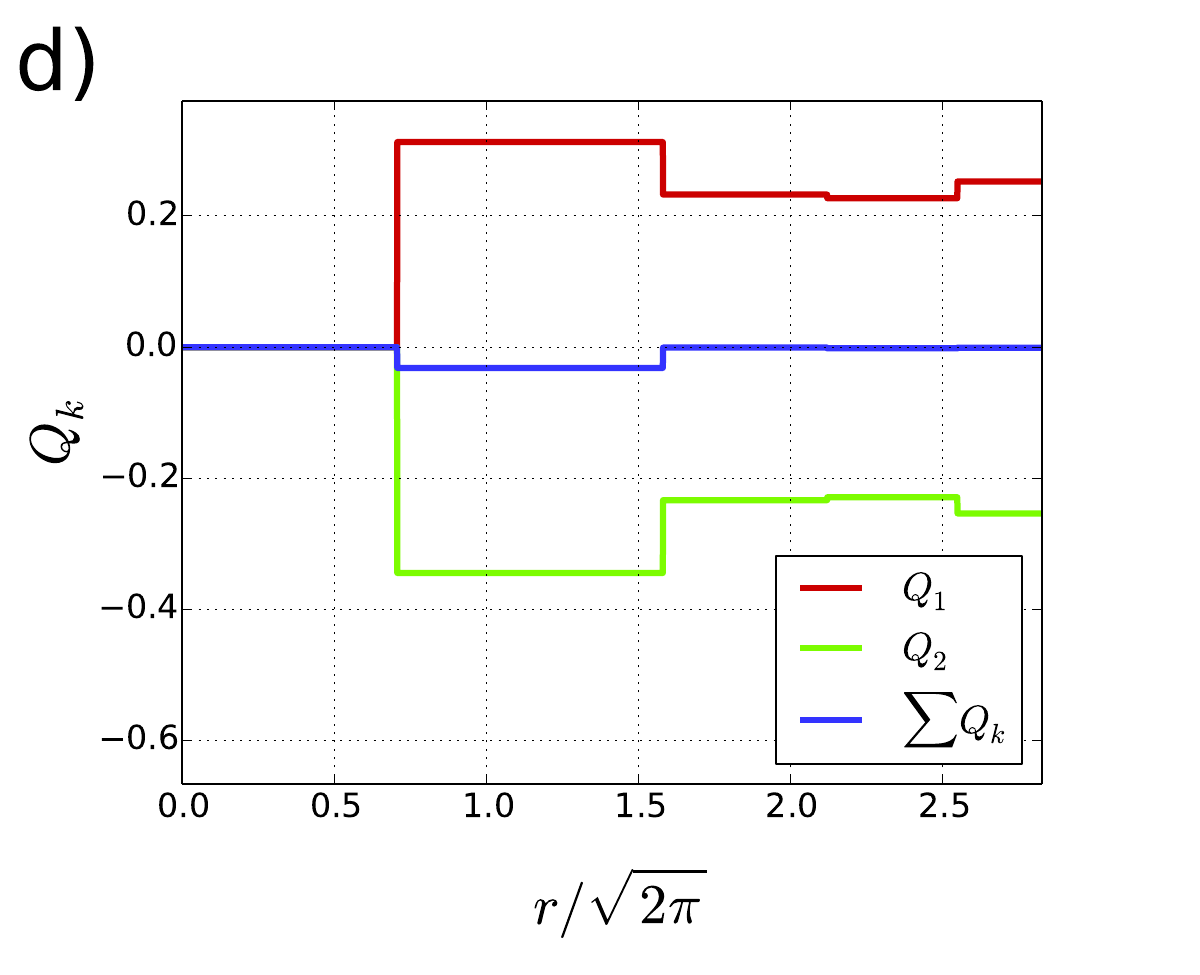}
        \includegraphics[width=0.245\textwidth]{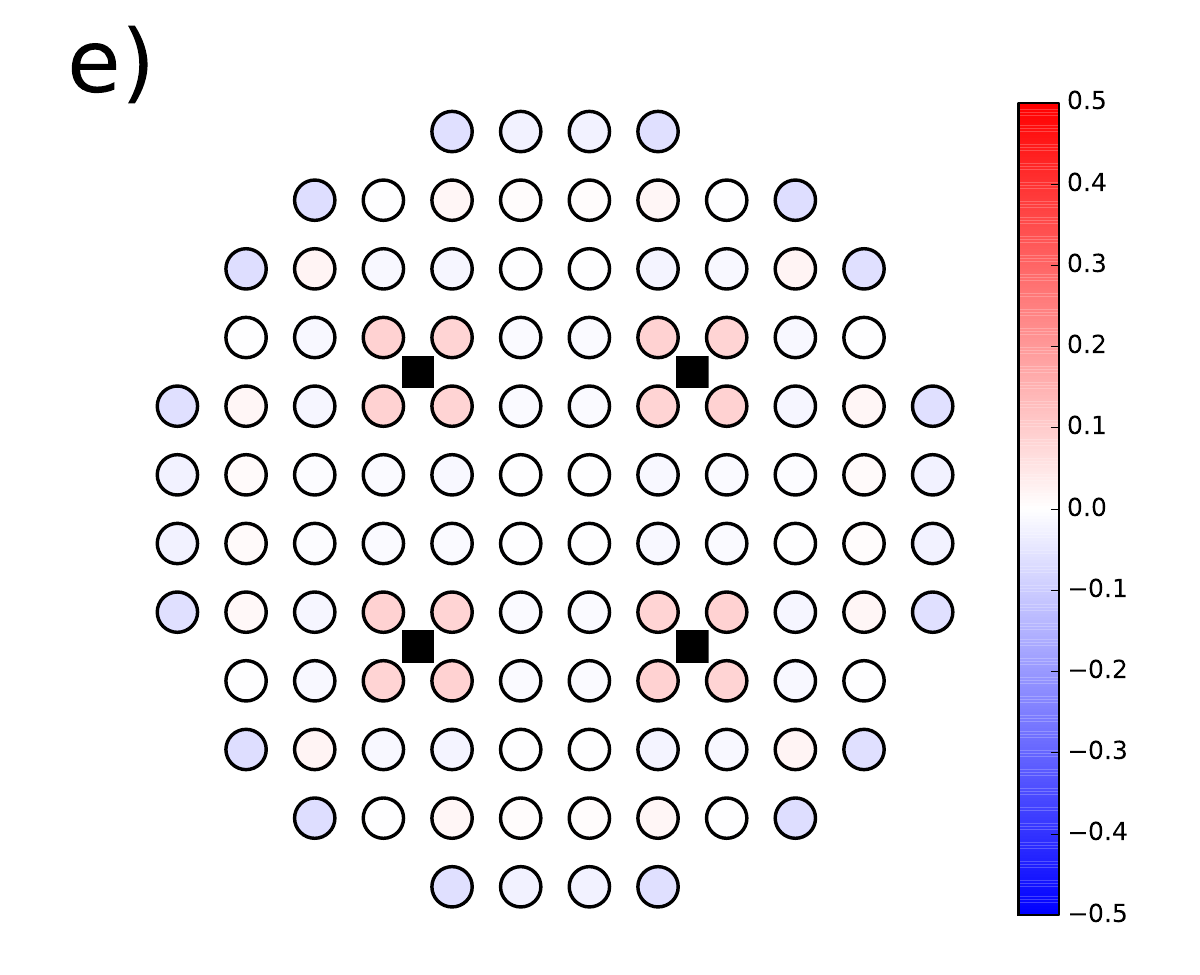}
        \includegraphics[width=0.245\textwidth]{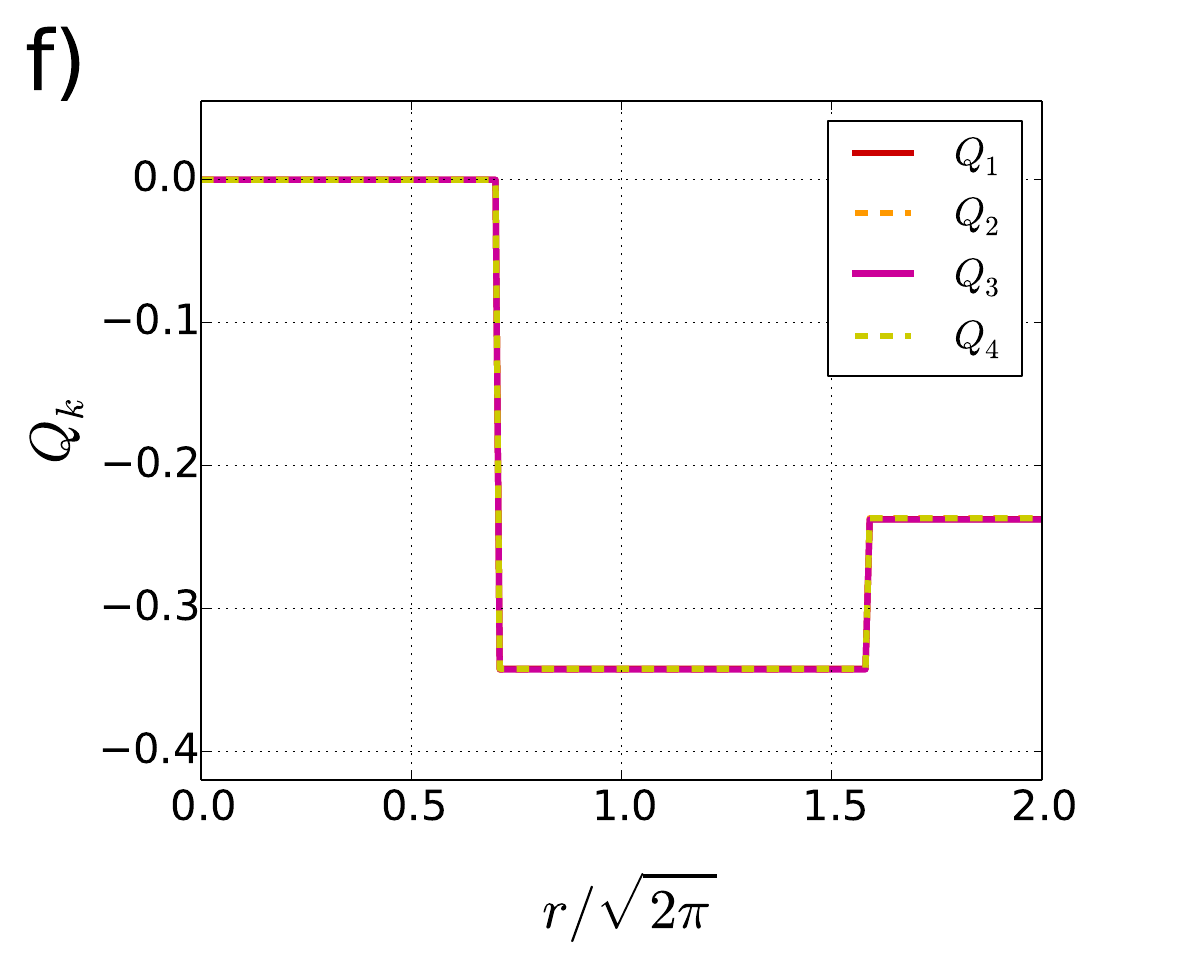}
        \includegraphics[width=0.245\textwidth]{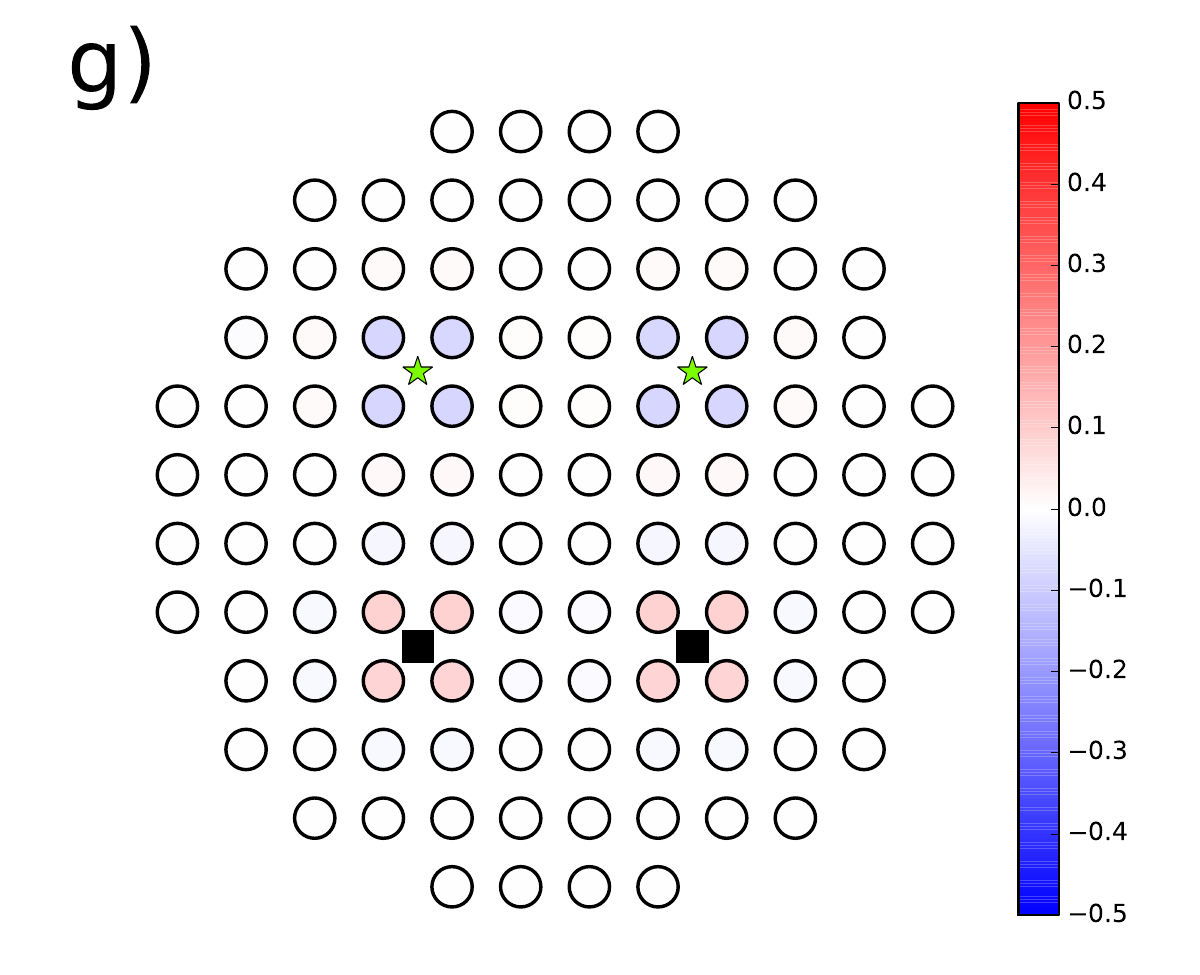}
        \includegraphics[width=0.245\textwidth]{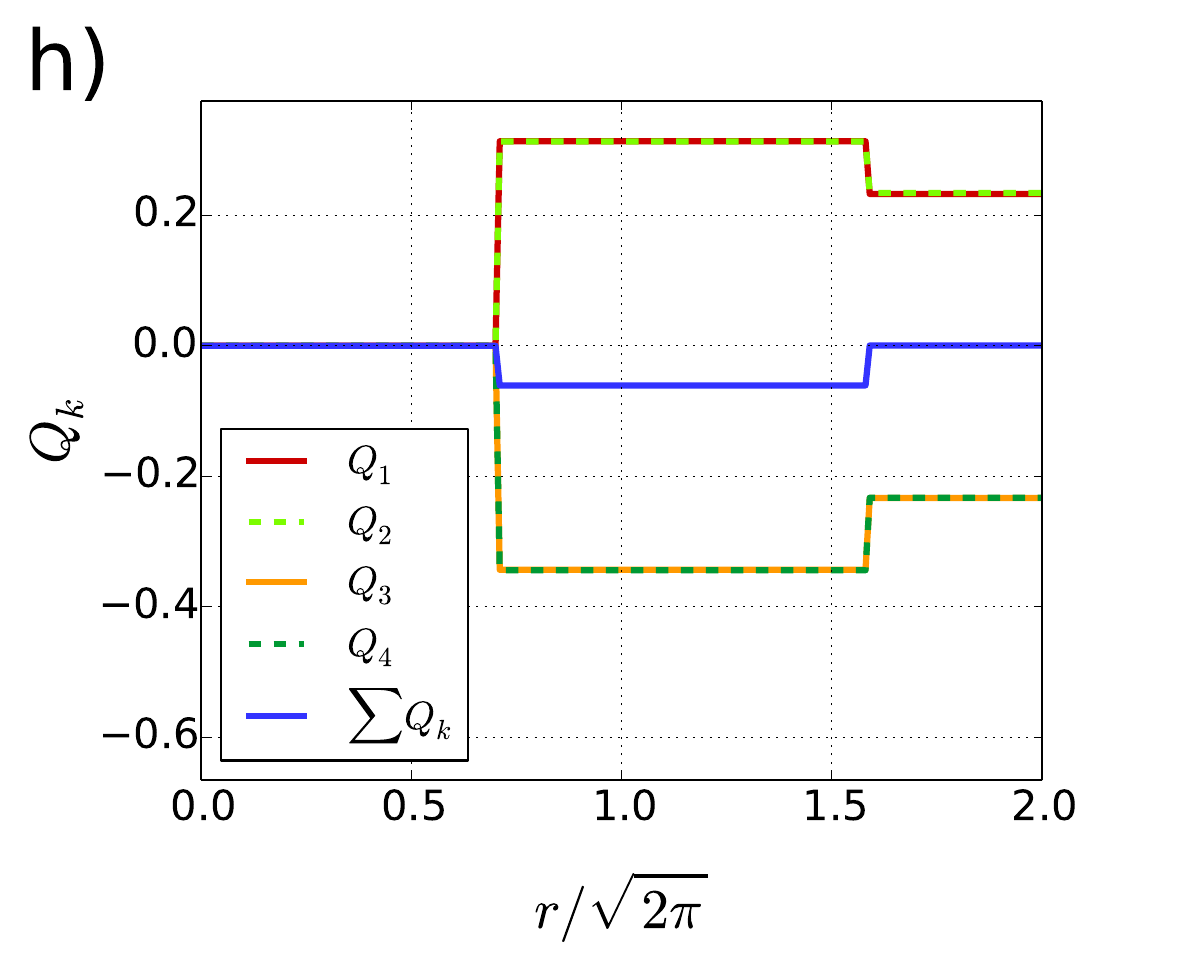}
	\caption{$a), c), e), g)$: Circles, stars and squares represent lattice sites, quasiholes and quasielectrons, respectively. We fix $N=112$. 
		The density profile $\rho(z_i)$
		defined as the particle density difference between the states with and without anyons, 
is plotted with colorbar for the cases of $a)$ two quasielectrons, $c)$ one quasihole-one quasielectron, $e)$ four quasielectrons and $g)$              
two quasiholes-two quasielectrons in the states, respectively. $b), d), f), h)$: The excess charge distributions are computed. The anyon charges approach $\sim \pm 0.25$ 
with an error of magnitude $10^{-4}$. 
Note that the density profiles are similar for quasiholes and quasielectrons except for a sign
as evident from the plot of the excess charge, $\sum_{k} Q_k$, of the anyons shown in $d)$ and $h)$, respectively.}\label{Fig_MR_anyon_density}
\end{figure*}

The state is defined as
\begin{equation}\label{wavefunc_1} 
|\Psi_\alpha\rangle = \frac{1}{C_\alpha}\sum_{n_1,....,n_N}\Psi_\alpha(\{z_j\},\{w_k\})|n_1,....,n_N\rangle
\end{equation} with 
\begin{equation}\label{wavefunc_norm} 
C_\alpha^2 = \sum_{n_1,....,n_N}|\Psi_\alpha(\{z_j\},\{w_k\})|^2
\end{equation}
 
 We insert an extra charge $\mathcal{P}/q$ at infinity to the state \eqref{wavefunc_1} by incorporating the operator $\Xi_\mathcal{P}(\infty) =   \       :e^{i\frac{\mathcal{P}}{\sqrt{q}}\phi(\infty)}:$ of charge $\frac{\mathcal{P}}{q}$, placed at infinity. Therefore we have the state with charge at infinity as
\begin{equation}\label{wavefunc_2} 
\Psi_\alpha(\{z_j\},\{w_k\}) \propto \langle 0| \Xi_\mathcal{P}(\infty) \prod_{k=1}^{Q}W_{p_k}(w_k)  \prod_{j=1}^{N}\mathcal{V}_{n_j}(z_j)|0\rangle
\end{equation} 
where $\langle 0|\cdots|0 \rangle$ is the vacuum expectation value in the CFT and $\alpha$ is a label that indicates which of the $2^{\frac{Q}{2}-1}$ conformal blocks is considered, 
where $Q$ is assumed to be even. As derived in Ref.\onlinecite{Anne8}, we write the states containing $Q$ quasiholes as follows 
\begin{equation}\label{wavefunc_3} 
\begin{split}
&\Psi_\alpha(\{z_j\},\{w_k\}) = \delta_n \langle \sigma(w_1)..\sigma(w_Q) \psi(z_1^\prime)..\psi(z_M^\prime) \rangle_\alpha
\\& \times \prod_{i<j}(z_{i}-z_{j})^{qn_{i}n_{j}}	
\prod_{i\neq j}(z_{i}-z_{j})^{-\eta n_{i}}	
\prod_{i<j}(w_i-w_j)^{p_ip_j/q}\\&
\times \prod_{i,j}(w_i-z_j)^{p_in_j}
\prod_{i,j}(w_i-z_j)^{-\eta p_i/q} 
\end{split}
\end{equation} 
where $\delta_n = 1$ iff the total number of particles fulfills 
\begin{equation}\label{particle_number}
M = \sum_{j=1}^{N} n_j = (\eta N - \mathcal{P} - \sum_{k=1}^{Q}p_k)/q\;,
\end{equation}
otherwise $\delta_n = 0$. In \eqref{wavefunc_3}, we denote the $M$ occupied sites by $z^\prime_1, \cdots, z^\prime_M$.
In the case of no quasiholes $(Q = 0)$ the correlator in \eqref{wavefunc_3} simplifies to 
\begin{equation}\label{wavefunc_no_qh} 
\langle \psi(z_1^\prime)..\psi(z_M^\prime) \rangle =  \text{Pf}\Big(\frac{1}{z_i^\prime-z_j^\prime}\Big),
\end{equation} where Pf is the Pfaffian. For two quasiholes $(Q = 2)$ we arrive at \cite{German1}
\begin{equation}\label{wavefunc_6} 
\begin{split}
&\langle \sigma(w_1)\sigma(w_2)\psi(z_1^\prime)..\psi(z_M^\prime) \rangle =   2^{-\frac{M}{2}} (w_1-w_2)^{-\frac{1}{8}}\\& \times \prod_{i,j}(w_i-{z^\prime_j})^{-\frac{1}{2}}\times \text{Pf}\bigg(\frac{(z_i^\prime - w_1)(z_j^\prime - w_2)+(i \longleftrightarrow j)}{(z_i^\prime-z_j^\prime)}\bigg)
\end{split}
\end{equation}and for four quasiholes $(Q = 4)$ we have \cite{German1}
\begin{equation}\label{wavefunc_4} 
\begin{split}
\langle \sigma(w_1)&\sigma(w_2)\sigma(w_3)\sigma(w_4)\psi(z_1^\prime)....\psi(z_M^\prime) \rangle_{\alpha} \\&= 2^{-\frac{M+1}{2}}
(w_1-w_2)^{-\frac{1}{8}} (w_3-w_4)^{-\frac{1}{8}}  \\& \times \prod_{i,j}(w_i-z_j^\prime)^{-\frac{1}{2}}
\bigg ((1-x)^{\frac{1}{4}} + \frac{(-1)^{\mathrm{m}_\alpha}}{(1-x)^{\frac{1}{4}}}\bigg )^{-\frac{1}{2}}  \\&
\times \bigg( (1-x)^{\frac{1}{4}} \Phi_{(13)(24)} + (-1)^{\mathrm{m}} (1-x)^{-\frac{1}{4}} \Phi_{(14)(23)} \bigg)
\end{split}
\end{equation} with
\begin{eqnarray*}
& &\Phi_{(k_1k_2)(k_3k_4)} = \\
& &\text{Pf}\bigg(  \frac{(w_{k_1}-z_i^\prime)(w_{k_2}-z_i^\prime)(w_{k_3}-z_j^\prime)(w_{k_4}-z_j^\prime) + (i \longleftrightarrow j)}{(z_i^\prime-z_j^\prime)} \bigg).\\
& &
\end{eqnarray*}
The parameter 
\begin{equation}
x = \frac{(w_1-w_2)(w_3-w_4)}{(w_1-w_4)(w_3-w_2)}
\end{equation} 
is the anharmonic ratio. The Pfaffian requires $M$ to be even in order to be non-zero. 
Depending on the non trivial fusion algebra $\sigma \times \sigma = I + \psi$ and the fusion channel we have 
two linearly independent degenerate states $\Psi_{I}$ and $\Psi_{\psi}$ (i.e.\, $\alpha \in \{I,\psi\}$) and we have $m_I = 0$ and $m_\psi = 1$ for the four quasiholes case.

{\it Here, we claim that Eq.\eqref{wavefunc_3}-Eq.\eqref{wavefunc_4} define quasielectron states if we take some or all values of $p_k$ to be negative.} 
We establish our statement by investigating the density profile, charge, shape and braiding statistics of the anyons in the states containing two quasielectrons, 
one quasihole-one quasielectron, four quasielectrons and two quasiholes-two quasielectrons on a square lattice. We set $q=2$ in all numerical computations throughout this section. 

\subsection{Density profile, charge and shape of the anyons}

For any state $\Phi$ we define $\langle n(z_i) \rangle = \langle \Phi | n(z_i) | \Phi \rangle$ as the lattice density of the $i$th lattice site. Therefore
\begin{equation}\label{Density_Profile}
\rho(z_i)=[\langle n(z_i) \rangle_{Q \neq 0} - \langle n(z_i) \rangle_{Q = 0}]
\end{equation}
gives the density profile where $\langle n(z_i) \rangle_{Q}$ is the particle density at the $i$th lattice site in the presence of $Q$ anyons. 
If we take the standard fermionic particle charge as $-1$ then the excess charge of the $k$th anyon is defined to be the sum of minus $\rho(z_i)$ over a circular region of radius $r$ around the anyon 
\begin{equation}\label{Excess_Charge}
Q_k = -\sum_{i} \rho(z_i), \ |z_i-w_k| \leq r\;.
\end{equation}
The charge of the anyon is the value that the excess charge converges to for large $r$ when the region is far from the edge and other anyons in the system.

We exploit Eq.\eqref{Density_Profile} and Eq.\eqref{Excess_Charge} to compute the density profile, charge and shape of the anyons in the system. Here, we respectively focus on four different 
system flavors. We investigate systems that respectively contain two and four quasielectrons, and systems that contain either one quasihole and one quasielectron or 
two quasiholes and two quasielectrons. We note from Eq.\eqref{particle_number} that the total number of particles $M$ can not in general be made even for both cases of the states 
containing anyons and without anyons if we take the same value for $\eta$ and put the charge at infinity to zero. We would like $\eta = 1$ both with and without anyons, and we hence choose a suitable $\mathcal{P}$ to make $M = N/q$ in all the cases. 

Fig \ref{Fig_MR_anyon_density}.$a), c), e),$ and $g)$ display the density profiles for systems containing two quasielectrons, one quasihole-one quasielectron, 
four quasielectrons and two quasiholes-two quasielectrons in the states respectively and we choose the values for $\mathcal{P}$ accordingly as $+1,0,+2,0$ in the states containing anyons. 
The state without anyons has $\mathcal{P}=0$ always. The non zero charges at infinity lead to the edge effects in the two quasielectrons and four quasielectrons cases 
(see Fig \ref{Fig_MR_anyon_density}.$a)$ and $e)$). 
We plot the excess charge distribution of the anyons as a function of the radial distance from the anyons as shown in 
Fig \ref{Fig_MR_anyon_density}.$b), d), f)$ and $h)$. We observe that the anyons are localized with radii of a few lattice constants. Moreover, they achieve 
a charge of $p_k/q \sim \pm 0.25$, while we take $p_k = \pm 1/2$, consistent with the quasihole charge in the continuum and as reported in experimental findings \cite{Others1,Others5}. 
This provides support for the claim that those anyons are of Ising anyon nature. 
Also, the density profiles are very similar for the quasiholes and quasielectrons except for the sign as evident from the plot of $\sum_{k} Q_k$ 
in Fig \ref{Fig_MR_anyon_density}.$c)$ and  $g)$. 

\subsection{Fractional braiding statistics of the anyons}

\begin{figure*}[!t]
        \centering
  \begin{tabular}[b]{cc}
          \begin{tabular}[b]{c}
                  \subfloat[]{
		  \includegraphics[width=0.48\textwidth]{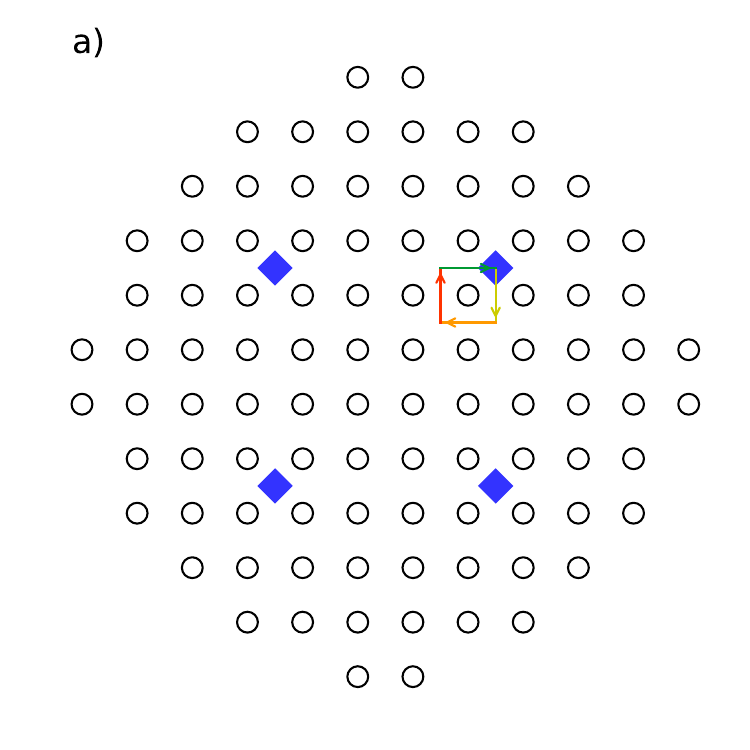}
                             }
    \\
    \textcolor{white}{.}\\
    \textcolor{white}{.}\\
          \end{tabular}
&
    \begin{tabular}[b]{c}
            \subfloat[] {
	    \includegraphics[width=0.48\textwidth]{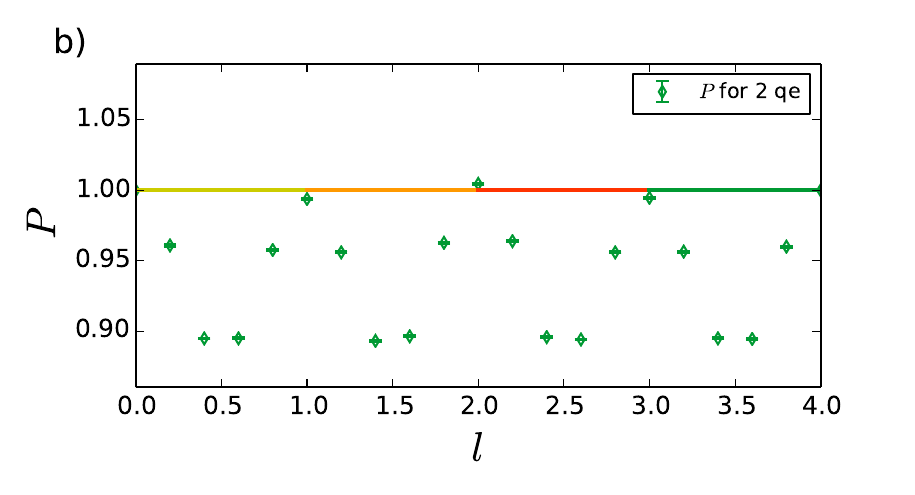}
                        }
      \\
      \subfloat[] {
      \includegraphics[width=0.48\textwidth]{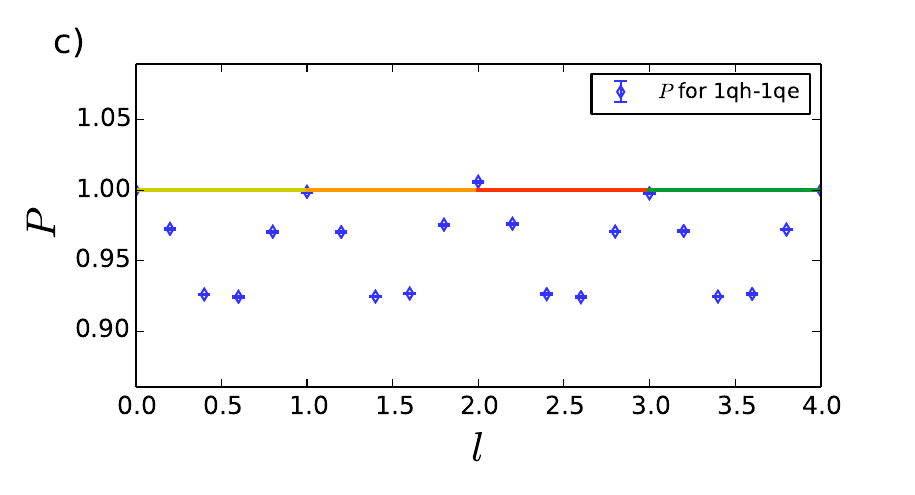}
                  }
    \end{tabular}
  \end{tabular}
  \caption{$a)$ Lattice sites and anyons are denoted by circles and diamonds, respectively. Proper choices of the charges of the anyons lead to the
          configurations shown in Fig \ref{Fig_MR_anyon_density}. The anyons are placed in the bulk and sufficiently separated from each other.
          Subsequently one anyon is moved around one lattice site through a closed loop along the path midway in the lattice plaquette while keeping the other anyon
          positions fixed. The lattice size is set to $N = 96$. $b)$ and $c)$ We take the set up as described in $a)$.
          We plot the overlap quantity $P$ as a function of different moves i.e. the position $l$
	  of the circulating anyon for the cases of two quasielectrons (2 qe) and one quasihole-one quasielectron (1qh-1qe) as shown in $b)$ and $c)$, respectively.
          For both cases, we note that the quantities vary with the period of the lattice up to finite size effects.
  } \label{Fig_MR_anyon_braiding_lattice_view}
\end{figure*}
While braiding, we adiabatically circulate one anyon at position $w_k$ around another anyon along a closed path. This transforms the state as 
\begin{equation}
|\Psi_\alpha\rangle \longrightarrow \gamma_A \gamma_M \gamma_B |\Psi_\alpha\rangle 
\end{equation}
where $\gamma_A, \gamma_M$ and $\gamma_B = e^{i \theta_B}$ 
are respectively the Aharonov-Bohm phase arising when a charged particle circulates in a background magnetic field, the monodromy matrix 
(phase matrix from the analytic continuation) and the Berry matrix with elements 
\begin{equation}\label{Berry_1}
\begin{split}
\big[\theta_B\big]_{\alpha\beta} =i\sum_{k=1}^Q\oint_\Gamma \Big(\langle\Psi_\alpha|\frac{\partial \Psi_\beta}{\partial w_k}\rangle dw_k
+\langle\Psi_\alpha|\frac{\partial \Psi_\beta}{\partial \bar{w}_k}\rangle d\bar{w}_k\Big)
\end{split}
\end{equation}  
where $\alpha,\beta \in \{I,\psi\}$. The states under consideration here are normalized. Following Ref.\onlinecite{Anne8}, if we show that the states are orthogonal, 
i.e. $\langle \Psi_\alpha | \Psi_\beta \rangle = \delta_{\alpha \beta}$, then we have
\begin{equation}\label{Berry_2a}
\begin{split}
\big[\theta_B\big]_{\alpha\beta}=i \delta_{\alpha\beta}  \oint_\Gamma I dw_k + c.c. 
\end{split}
\end{equation}  
with 
\begin{equation}
I =  \frac{1}{C_\alpha} \Big( \frac{\partial C_\alpha}{\partial w_k} \Big) = \frac{\partial ln(C_\alpha)}{\partial w_k}\,.
\end{equation}
Therefore, if $C_\alpha$ is periodic 
in $w_k$, then $\big[ \gamma_B \big]_{\alpha \beta} = \delta_{\alpha \beta}$ and the braiding statistics would be given by $\gamma_M$ alone. 
Also, the braiding properties would be the same as for the continuum if we show that $C_\alpha = C_\beta$. Hence we write down the following two sufficient conditions as follows 
\begin{enumerate}
	\item[(i)]$|\sum_{n_i} \Psi^*_\alpha \Psi_{\beta}| = \mathcal{C} \delta_{\alpha \beta} $ up to exponentially small finite size effects and $\mathcal{C}$ is a constant, and 
	\item[(ii)]  $C_\alpha$ is periodic when we move one anyon through a closed loop.  
\end{enumerate}
We now proceed by explicitly computing fractional braiding statistics with two and four anyons in the system. We employ  Metropolis Monte Carlo
simulations to calculate the desired properties of the states.\\

{\it Two anyons scenario:} In this scenario, we incorporate either two quasielectrons or one quasihole and one quasielectron in the state. 
Since we have only one state in this case, condition (ii) is the only condition to be satisfied. We place the anyons in the bulk and isolated 
from each other and move one anyon around one lattice site through a closed loop while keeping the other anyon fixed. 
We choose the path to be along the midway in the lattice plaquette as shown in Fig \ref{Fig_MR_anyon_braiding_lattice_view}.$a)$ 
and expect the same outcome to hold if we choose any other closed path as well. Proper choice of the charges of the anyons in 
Fig \ref{Fig_MR_anyon_braiding_lattice_view}.$a)$ leads to the configurations as depicted in Fig \ref{Fig_MR_anyon_density}.$a)$ and $c)$, respectively. 
We define the following quantity to investigate
\begin{equation}\label{Berry_3}
\begin{split}
P = \frac{C_0^2}{C_l^2}
\end{split}
\end{equation} 
where $C_{l}$ and $C_{0}$ denote the normalization constants when the moving anyon is at position $l$ and its initial position $l = 0$, respectively. Fig \ref{Fig_MR_anyon_braiding_lattice_view}.$b)$ and $c)$ 
show the above mentioned quantity $P$ as a function of the different moves $(l)$ of the moving anyon respectively for the cases of two quasielectrons and 
one quasihole-one quasielectron in the state. We find the periodic variations of the quantity $P$ with different positions of the circulating anyon 
(upto finite size effects). 
\begin{figure*}
                \includegraphics[width=0.370\textwidth]{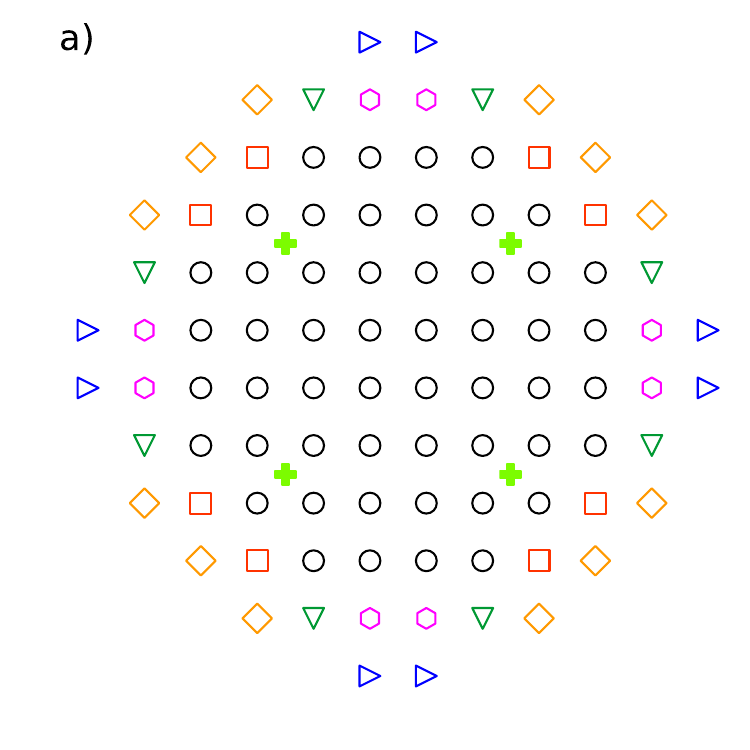}\;\;\;\;\;\;\;\;\;\;\;\;\;\;\;\;\;\;\;\;\;\;\;\;\;\;\;
		\includegraphics[width=0.485\textwidth]{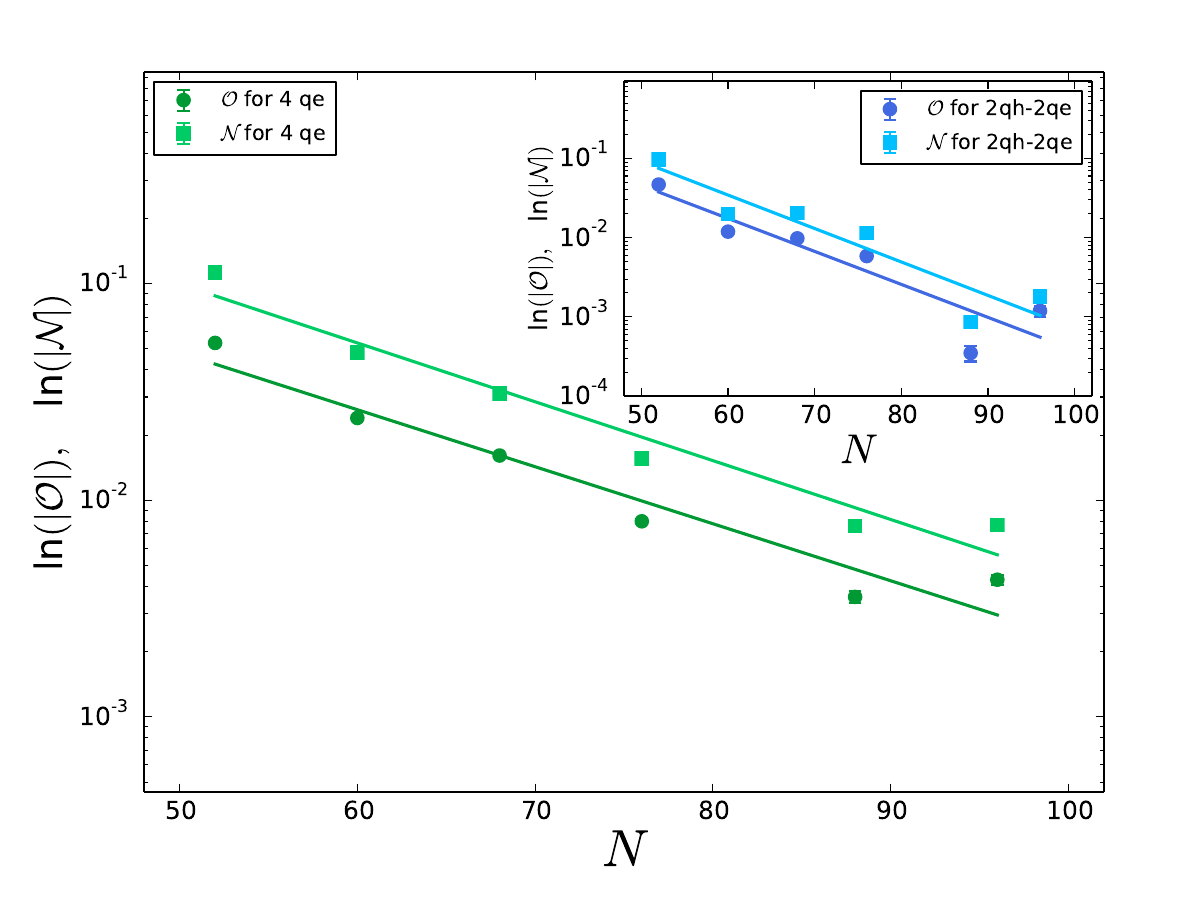}
\caption{We keep the anyons (pluses) fixed in the bulk and sufficiently separated from each other and increase the lattice size $N$ by adding
more sites as shown in $a)$. We denote different lattice sizes with symbols as $N = 52$ (circles), $N = 60$ (squares added), $N = 68$ (hexagons added), $N = 76$ (down pointing triangles added),
$N = 88$ (diamonds added) and $N = 96$ (right pointing triangles added). Proper choices of the charges of anyons lead to the configurations as shown in Fig \ref{Fig_MR_anyon_density}.$e)$
and  $g)$, respectively.
In $b)$ we plot the overlaps $\mathcal{O}$ (circles) and $\mathcal{N}$ (squares) as defined in \eqref{Berry_2b} and \eqref{Berry_2c}, respectively, in the semi log scale
as a function of the lattice size for the cases of four quasielectrons (4 qe) and two quasiholes-two quasielectrons (2qh-2qe) present in the system.
We note that the quantities of interest in both the plots are following an exponential decay for sufficiently large lattice sizes. A linear fitting indeed proves the
exponential decay implying that the states $\Psi_{I}$ and $\Psi_{\psi}$ are orthogonal in the thermodynamic limit.
}\label{Fig_MR_non_Abelian_anyon_overlap}
\end{figure*}		
Therefore, we find the Berry phase contribution as $\gamma_B = 1$. Hence, the braiding statistics are given by the monodromy matrix only and under 
the exchange of $w_j$ and $w_k$ we have 
\begin{equation}\label{Berry_3666}
\gamma_M =  e^{i \pi \big[ \frac{p_jp_k}{q} - \frac{1}{8} \big]}
\end{equation}
as a phase. 
Also, when an anyon circulates around the lattice sites in counter-clockwise fashion, we have the phase as 
\begin{equation}\label{Berry_3777}
\gamma_A = e^{-2 \pi i p_k /q}\;.
\end{equation}
This is interpreted as the Aharonov-Bohm phase of a particle with charge $p_k/q$, circulating around a closed loop and enclosing the background magnetic flux.\\

{\it Four anyons scenario:} We stress that in this case condition $(i)$ needs to be satisfied since we are now dealing with a two-fold degenerate ground state manifold. 
In order to investigate condition $(i)$ we study two quantities, namely the overlap of the two states 
\begin{equation}\label{Berry_2b}
\mathcal{O} =  \frac{| \sum_{n_i}  \Psi_I^{*}  \Psi_\psi| }{\sqrt{\sum_{n_i} |\Psi_I|^2  \sum_{n_i} |\Psi_\psi|^2}}
\end{equation}
and the ratio of the respective norms of the two states
\begin{equation}\label{Berry_2c}
\mathcal{N} = 1 - \frac{\sum_{n_i} |\Psi_I|^2}{\sum_{n_i} |\Psi_\psi|^2}\;.
\end{equation}
Again as in the previous two anyon case, we keep the four anyons fixed and sufficiently separated from each other and increase the lattice size by putting more lattice 
sites as shown in Fig \ref{Fig_MR_non_Abelian_anyon_overlap}.$a)$. A proper choice of the charges of the anyons in Fig \ref{Fig_MR_non_Abelian_anyon_overlap}.$a)$ 
leads to the configurations as in Fig \ref{Fig_MR_anyon_density}.$e), g)$. We proceed by plotting $\mathcal{O}$ and $\mathcal{N}$ in the semi log scale as a function 
of the system size in Fig \ref{Fig_MR_non_Abelian_anyon_overlap}.$b)$ 
for the cases of four quasielectrons (Fig \ref{Fig_MR_non_Abelian_anyon_overlap}.$b)$ main panel) and two quasiholes-two quasielectrons 
(Fig \ref{Fig_MR_non_Abelian_anyon_overlap}.$b)$ inset). We find an exponential decay of the quantities as a function of the system size. 
A linear fit determines the decay as $e^{-\lambda N}$ with the decay coefficient $\lambda$. For the case of four quasielectrons, the decay coefficient is $\lambda = 0.0263$ for $\mathcal{O}$ and $\lambda = 0.0271$ for $\mathcal{N}$. For the case of two quasiholes and two quasielectrons, the decay coefficient is $\lambda = 0.0416$ for $\mathcal{O}$ and $\lambda = 0.0422$ for $\mathcal{N}$. We show the linear fittings in 
Fig \ref{Fig_MR_non_Abelian_anyon_overlap}$b)$. 
Thus, in the thermodynamic limit $N \longrightarrow \infty$, the states are orthogonal with the same norm.
Therefore, we conclude that 
\begin{equation}
|\sum_{n_i} \Psi^*_\alpha \Psi_{\beta}| = \mathcal{C} \delta_{\alpha \beta} + \mathcal{O}(e^{-\lambda N})
\end{equation} 
where $\mathcal{C}$ is a constant and $\mathcal{O}(e^{-\lambda N})$ is a contribution which exponentially decays with the system size. 
 
\begin{figure*}
        \includegraphics[width=0.495\textwidth]{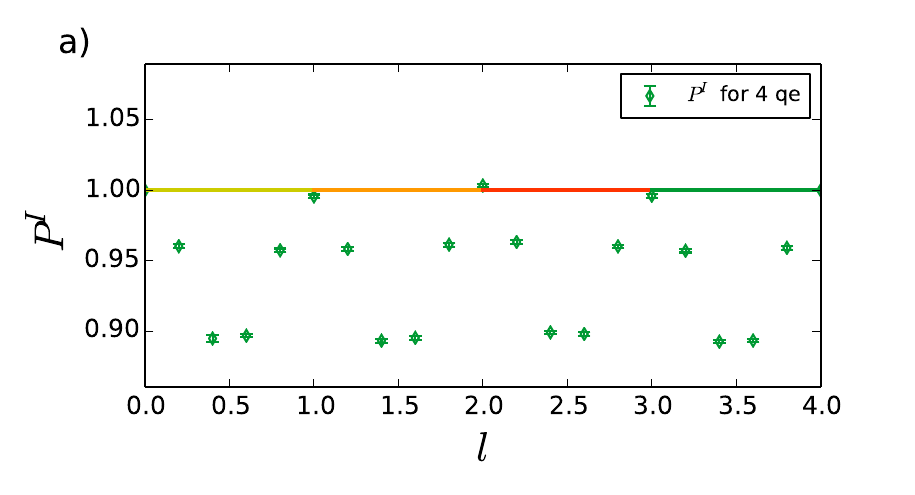}
        \includegraphics[width=0.495\textwidth]{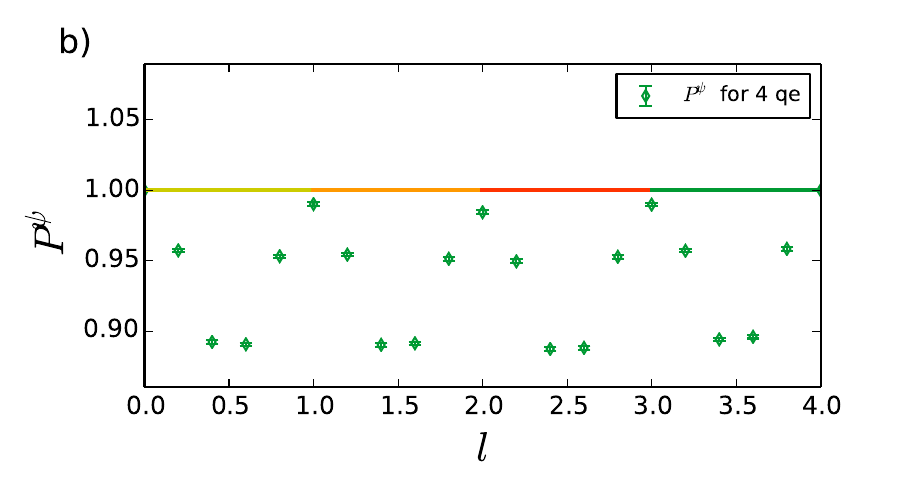}
        \includegraphics[width=0.495\textwidth]{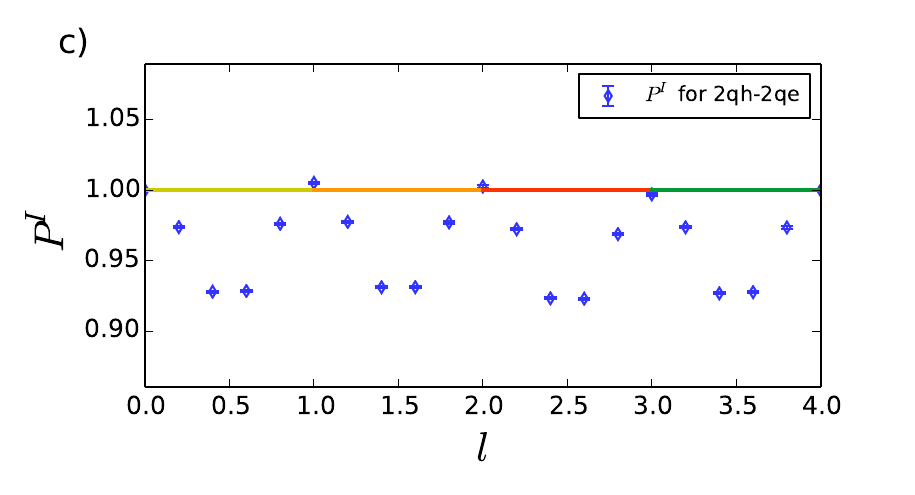}
        \includegraphics[width=0.495\textwidth]{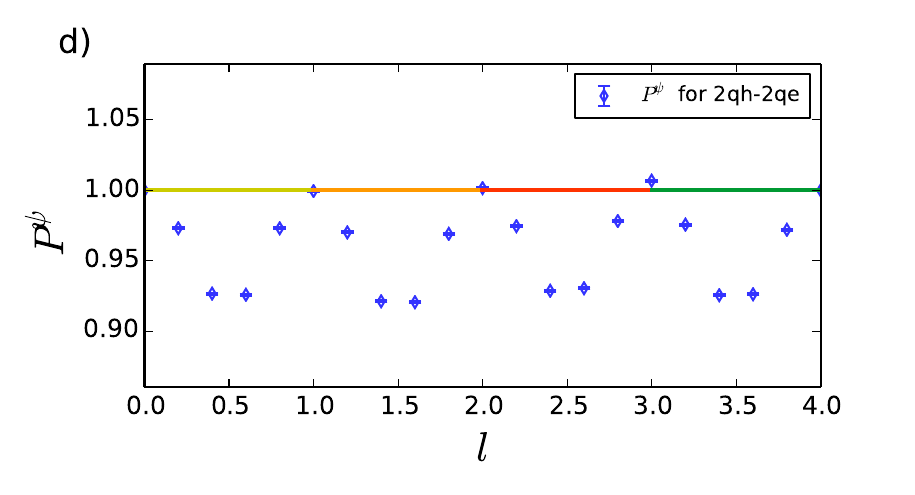}
	\caption{We take the set up as shown in Fig \ref{Fig_MR_anyon_braiding_lattice_view}.$a)$ and a proper choice of anyon charges leads to the configurations in
        Fig \ref{Fig_MR_anyon_density}.$e)$ and $g)$. We plot the overlap quantity $P^\alpha$ with $\alpha \in \{I,\psi\}$ as a function of different moves
	i.e. $l$ of the circulating anyon. In $a), b)$ we plot respectively $P^I, P^\psi$ for the case of four quasielectrons (4 qe) and in $c), d)$ concurrently we plot $P^I, P^\psi$ for the case of two quasiholes-two quasielectrons (2qh-2qe).
        It is seen that the quantities vary with the period of the lattice up to finite size effects.
}\label{Fig_MR_non_Abelian_anyon_braiding}
\end{figure*}

Henceforth, the overlap matrix becomes the identity matrix. Now to check the condition $(ii)$, we again place the anyons in the bulk and isolate them from each other 
and move one anyon around one lattice site through a closed loop while keeping the other anyons fixed. 
We choose the path to be along the midway in the lattice plaquette as shown in Fig \ref{Fig_MR_anyon_braiding_lattice_view}.$a)$ and expect the same outcome to hold if we 
choose any other closed path as well. Proper choice of the charges of the anyons in Fig \ref{Fig_MR_anyon_braiding_lattice_view}.$a)$ leads to the configurations as in 
Fig \ref{Fig_MR_anyon_density}.$e)$ and $g)$.  
Since the overlap matrix becomes diagonal for sufficiently large $N$, it is enough to investigate only the diagonal elements i.e.\ $\alpha = \beta$. 
We define the quantity to investigate as follows 
\begin{equation}\label{Berry_3a}
\begin{split}
P^\alpha = \frac{C_{\alpha 0}^2}{C_{\alpha l}^2}, \ \alpha \in \{I,\psi\} 
\end{split}
\end{equation}
$C_{\alpha l}$ and $C_{\alpha 0}$ denote the normalization constants when the moving anyon is at position $l$ and its initial position $l = 0$, respectively.
Fig \ref{Fig_MR_non_Abelian_anyon_braiding}.$a), b)$ and 
\ref{Fig_MR_non_Abelian_anyon_braiding}.$c), d)$ show the above mentioned quantity $P^{\alpha}$ as a function of the different moves $(l)$ 
of the moving anyon respectively for the cases of four quasielectrons and of two quasiholes-two quasielectrons. 
We again find the periodic variations of the quantity with different positions of the circulating anyon. 

Therefore, we satisfy both conditions $(i)$ and $(ii)$ and inscribe the Berry phase contribution as $\gamma_B = \hat{I}$ 
where $\hat{I}$ is the identity matrix. Hence, the braiding statistics are given by the monodromy matrices 
and under the exchange of $w_j$ and $w_k$ we inscribe 
\begin{equation}
[\Psi_I,\Psi_\psi]^T \mapsto \gamma_M^{j \rightleftarrows k} [\Psi_I,\Psi_\psi]^T\;.
\end{equation}
Now different choices of $j,k$ provide different monodromy matrices as
\begin{eqnarray}
\begin{split}	
\gamma_M^{1 \rightleftarrows 2 / 3 \rightleftarrows 4} &=&  e^{i \pi \big[ \frac{p_jp_k}{q} - \frac{1}{8} \big]} 
\begin{bmatrix}
1&0\\
0&i
\end{bmatrix}\;,\\
\gamma_M^{2 \rightleftarrows 3 / 1 \rightleftarrows 4} &=&  e^{i \pi \big[ \frac{p_jp_k}{q} + \frac{1}{8} \big]} \frac{1}{\sqrt{2}}
\begin{bmatrix}
1&-i\\
-i&1
\end{bmatrix}\;,\\ 
\gamma_M^{1 \rightleftarrows 3 / 2 \rightleftarrows 4} &=&  e^{i \pi \big[ \frac{p_jp_k}{q} + \frac{1}{8} \big]} \frac{1}{\sqrt{2}}
\begin{bmatrix}
1&-1\\
1&1
\end{bmatrix}
\end{split}
\end{eqnarray}
where we denote the exchange of the two anyons by the symbol $\rightleftarrows$. 
Also, when an anyon circulates around the lattice sites in counter-clockwise fashion, we have the phase as $\gamma_A = e^{-2 \pi i p_k /q}$. 
This is interpreted as the  Aharonov-Bohm phase of a particle with charge $p_k/q$, circulating around a closed loop and enclosing the background magnetic flux. 
As the monodromy matrices serve as the members of the Braid group, therefore we infer that the anyons are non-Abelian.

\subsection{Parent Hamiltonians}
We proceed by deriving a parent Hamiltonian for which \eqref{wavefunc_1} is the ground state. We exploit the results obtained in Ref.\onlinecite{Anne8} for quasiholes in the states 
(let us denote the state as $|\Psi_\alpha^{qh}\rangle$ with qh standing for quasiholes) and derive the parent Hamiltonian for our cases $(|\Psi_\alpha\rangle)$. 
We have the number of particles for $|\Psi_\alpha^{qh}\rangle$ as $M^{qh} = (\eta_{qh} N - \sum_{k}p_k^{qh})/q$ and the number of particles for $|\Psi_\alpha\rangle$ 
as $M = (\eta N - \sum_{k}p_k)/q$. Here, we keep the number of lattice sites $N$ fixed and $\eta_{qh}$ is for the state with quasiholes and $p_k^{qh}$ is the charge of 
the $k$th quasihole. Now to achieve the same number of particles in both the states we must have $M^{qh} = M$. This gives rise to the following condition
\begin{equation}\label{eta_cond}
\begin{split}
\eta = \eta_{qh} - \frac{1}{N} \big(  \sum_{k}p_k^{qh} - \sum_{k}p_k  \big)\,.
\end{split}
\end{equation}

 We note that $T|\Psi_\alpha\rangle = |\Psi_\alpha^{qh}\rangle$ where 
\begin{equation}\label{Ham_3}
\begin{split}
T = \prod_{i}\beta_i^{n_i}\theta_i^{n_i}
\end{split}
\end{equation}
 with 
 \begin{equation}\label{Ham_4}
 \begin{split}
 \beta_i = \prod_{j \neq i}(z_i-z_j)^{(\eta-\eta_{qh})}
 \end{split}
 \end{equation}
 and 
  \begin{equation}\label{Ham_5}
 \begin{split}
 \theta_i = \prod_{j}(w_j-z_i)^{(p_j^{qh}-p_j)}\,.
 \end{split}
 \end{equation}

For the rest of the section, we set $\eta_{qh} = 1$. In Ref.\onlinecite{Anne8}, it was shown that a set of operators $\lambda^a_i, \ a\in \{0,..,q-1\}$ annihilate the state $|\Psi_\alpha^{qh}\rangle$, i.e.\
 \begin{equation}
 \lambda^a_i |\Psi_\alpha^{qh} \rangle = 0\,.
 \end{equation}
We can re-write this as 
\begin{equation}
T^{-1}\lambda^a_i T |\Psi_\alpha \rangle = 0\,.
\end{equation}
Therefore, we define $\Lambda^a_i = T^{-1}\lambda^a_i T$, and it follows that $\Lambda^a_i |\Psi_\alpha \rangle = 0$. Utilizing the expressions for $\lambda^a_i$ from Ref.\onlinecite{Anne8}, we find
\begin{equation}\label{Ham_1} 
\begin{split}
\Lambda^{0} &=\sum_{i} \beta_i\theta_i d_i\, , 
\end{split}
\end{equation} 
\begin{equation}\label{Ham_6} 
\begin{split}
\underset{p=1,..,q-2}{\Lambda_i^{p}} =\sum_{j(\neq i)} \frac{\beta_j\theta_j d_j n_i}{(z_i-z_j)^{p}},\; 
\end{split}
\end{equation} 
\begin{equation}\label{Ham_2} 
\begin{split}
\Lambda_i^{q-1} = &\sum_{j(\neq i)} \frac{\beta_j\theta_j d_j n_i}{(z_i-z_j)^{q}}  + \sum_{j(\neq i)} \sum_{h(\neq i)}  \frac{[qn_j-1]\beta_h\theta_h d_hn_i}{(z_i-z_h)^{q-1}(z_i-z_j)}\\&
+  \sum_{j} \sum_{h(\neq i)} \frac{p_j\beta_h\theta_h d_hn_i}{(z_i-z_h)^{q-1}(z_i-w_j)}
\end{split}
\end{equation}  
where $d_j$ is the hardcore 
boson (fermion) annihilation operator at the $j$th lattice site for $q$ odd (even) and $n_j = d^\dagger_j d_j$. Therefore we write the positive semidefinite operator 
\begin{equation}\label{Ham_7} 
\begin{split}
H=\sum_{i=1}^N \sum_{a=0}^{q-1} \Lambda_i^{a\dagger}\Lambda_i^{a}
\end{split}
\end{equation}
as the parent Hamiltonian of \eqref{wavefunc_1}. The ground state of the Hamiltonian is degenerate for the cases of four quasielectrons and two quasiholes-two quasielectrons present in the 
systen which also signifies the non-Abelian nature of the anyons in our system. The Hamiltonian is long ranged and contains five-body interactions. Besides being an exact Hamiltonian for the anyons, it could be useful for testing numerical techniques. Also, it may be a starting point to find simpler Hamiltonians with the same ground state physics \cite{Anne4, Others50}.

\begin{figure*}
	\includegraphics[width=0.245\textwidth]{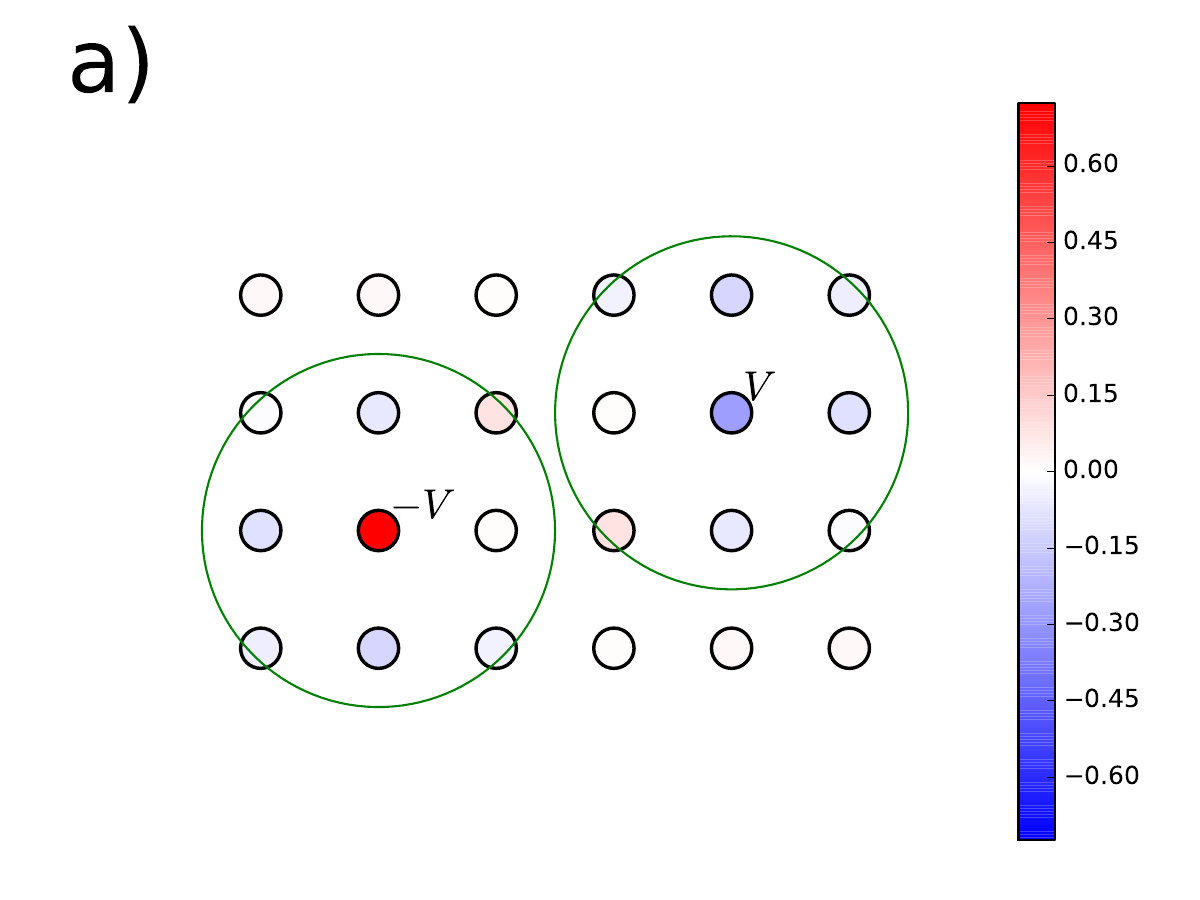}
	\includegraphics[width=0.245\textwidth]{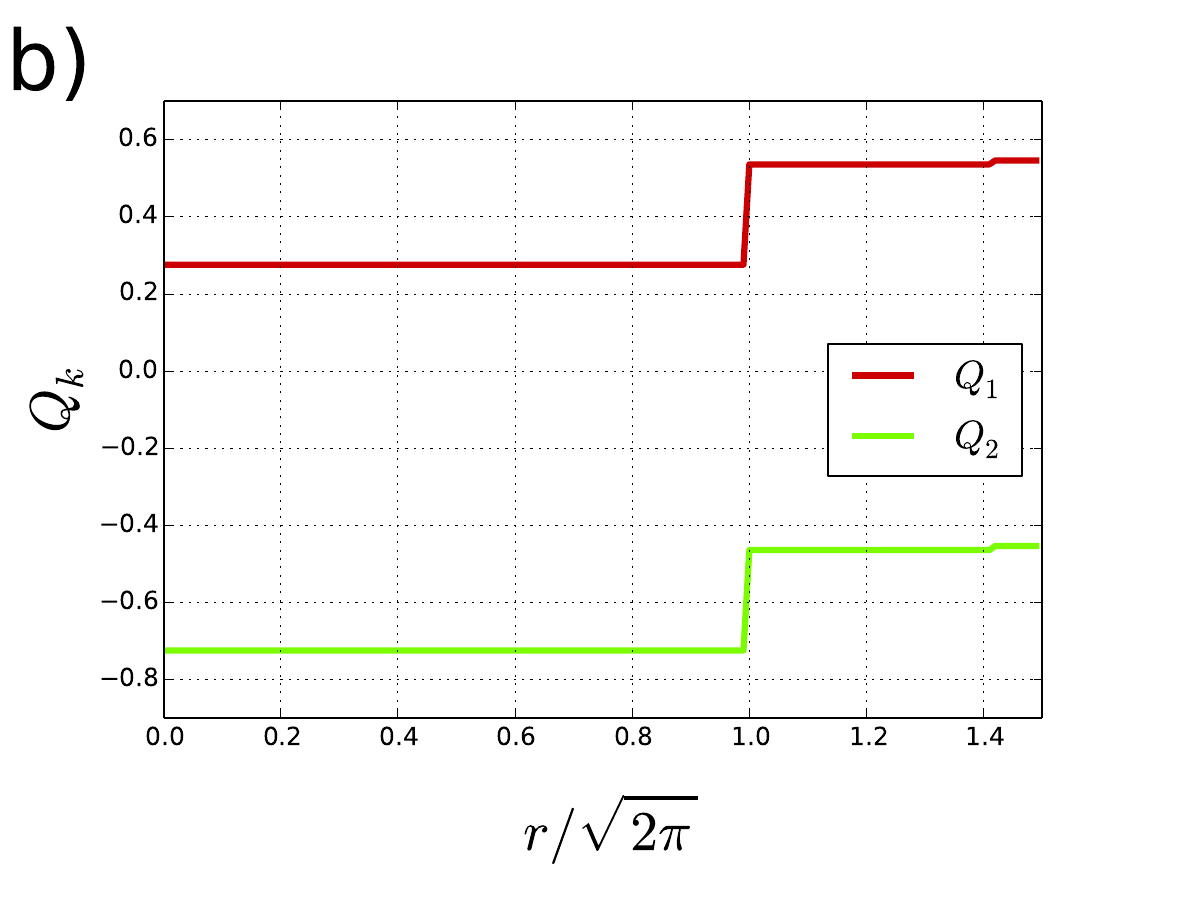}
	\includegraphics[width=0.245\textwidth]{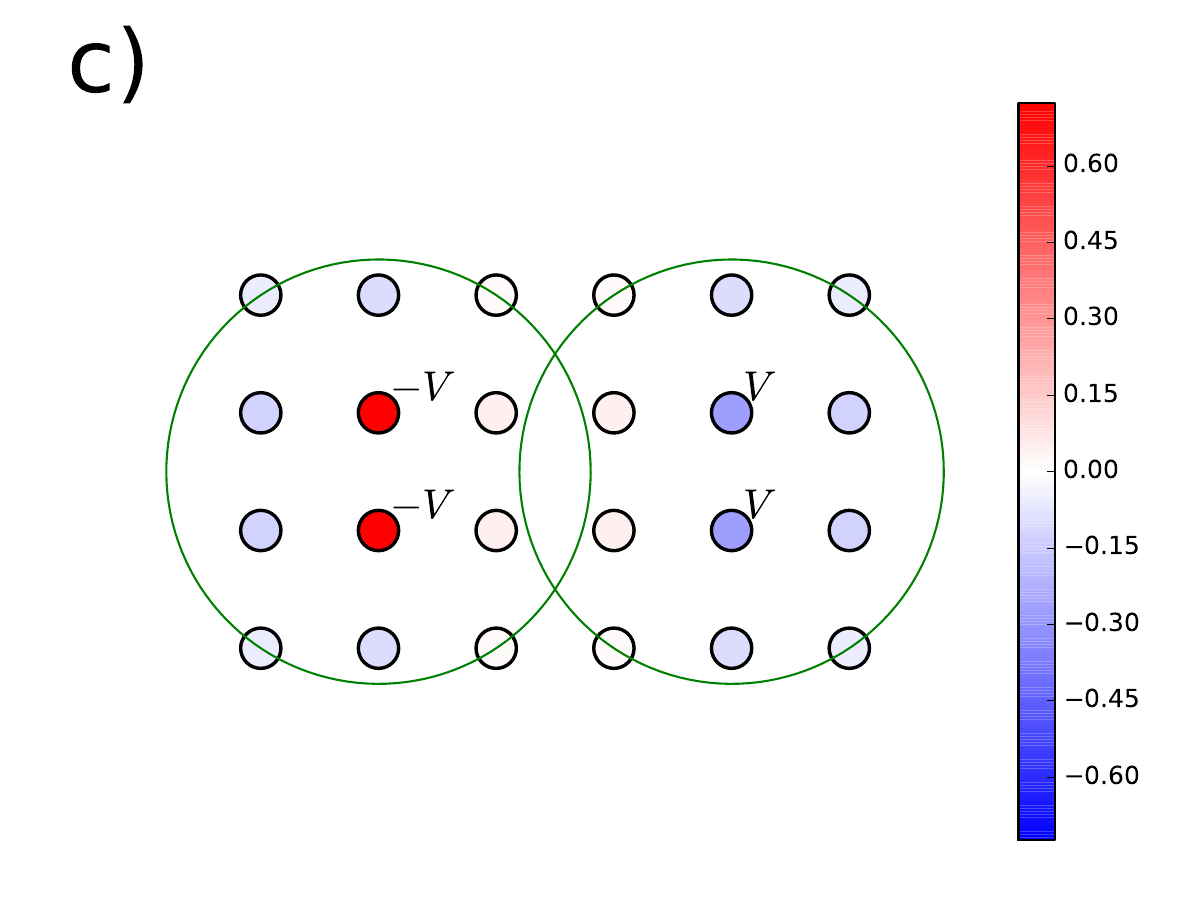}
	\includegraphics[width=0.245\textwidth]{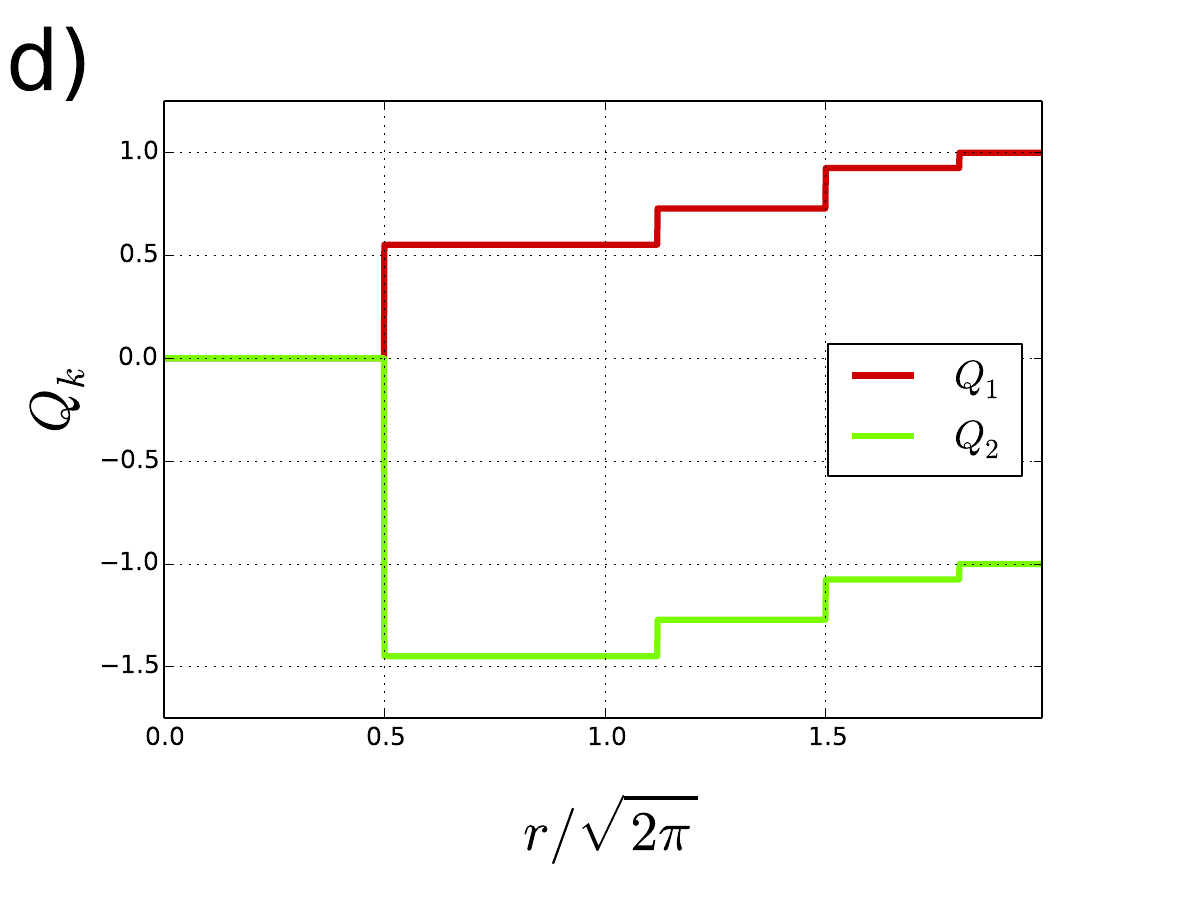}
	\includegraphics[width=0.245\textwidth]{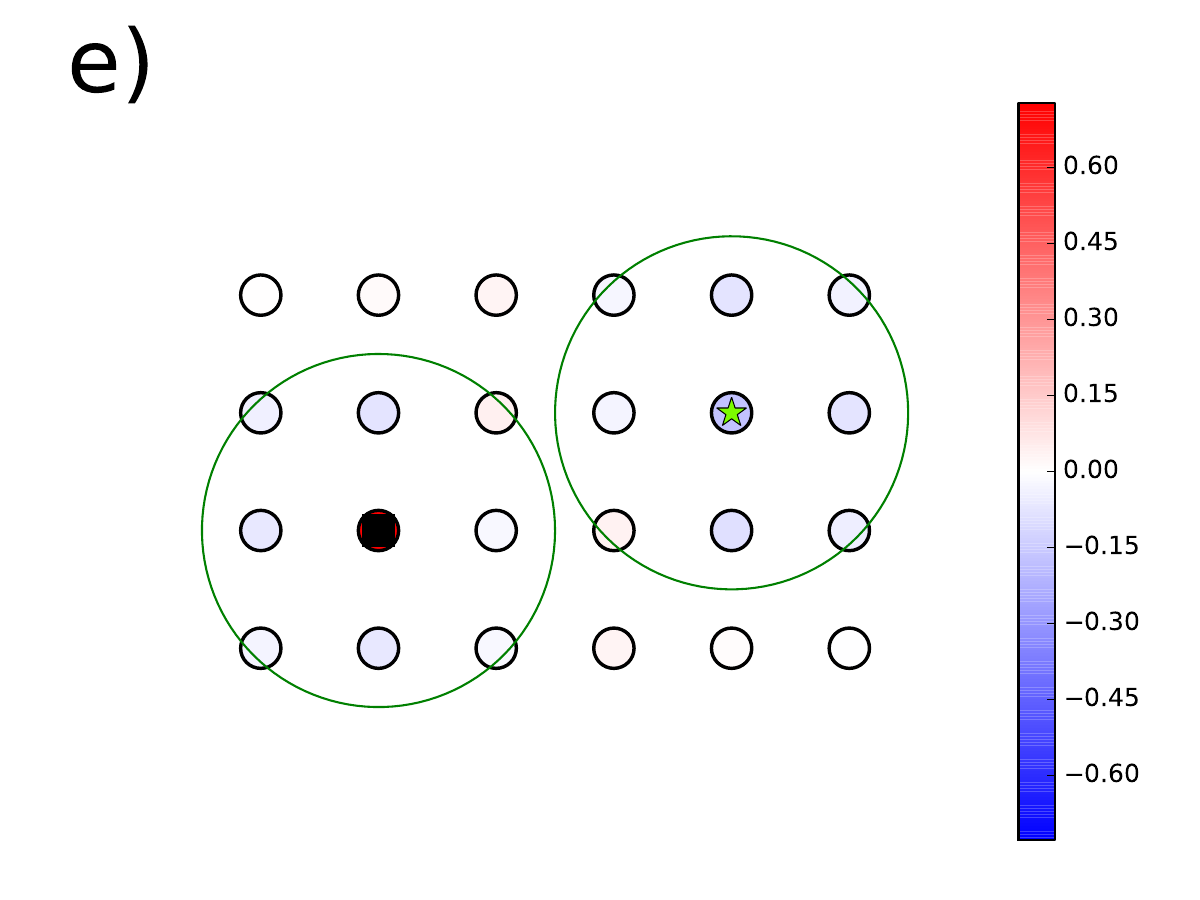}
	\includegraphics[width=0.245\textwidth]{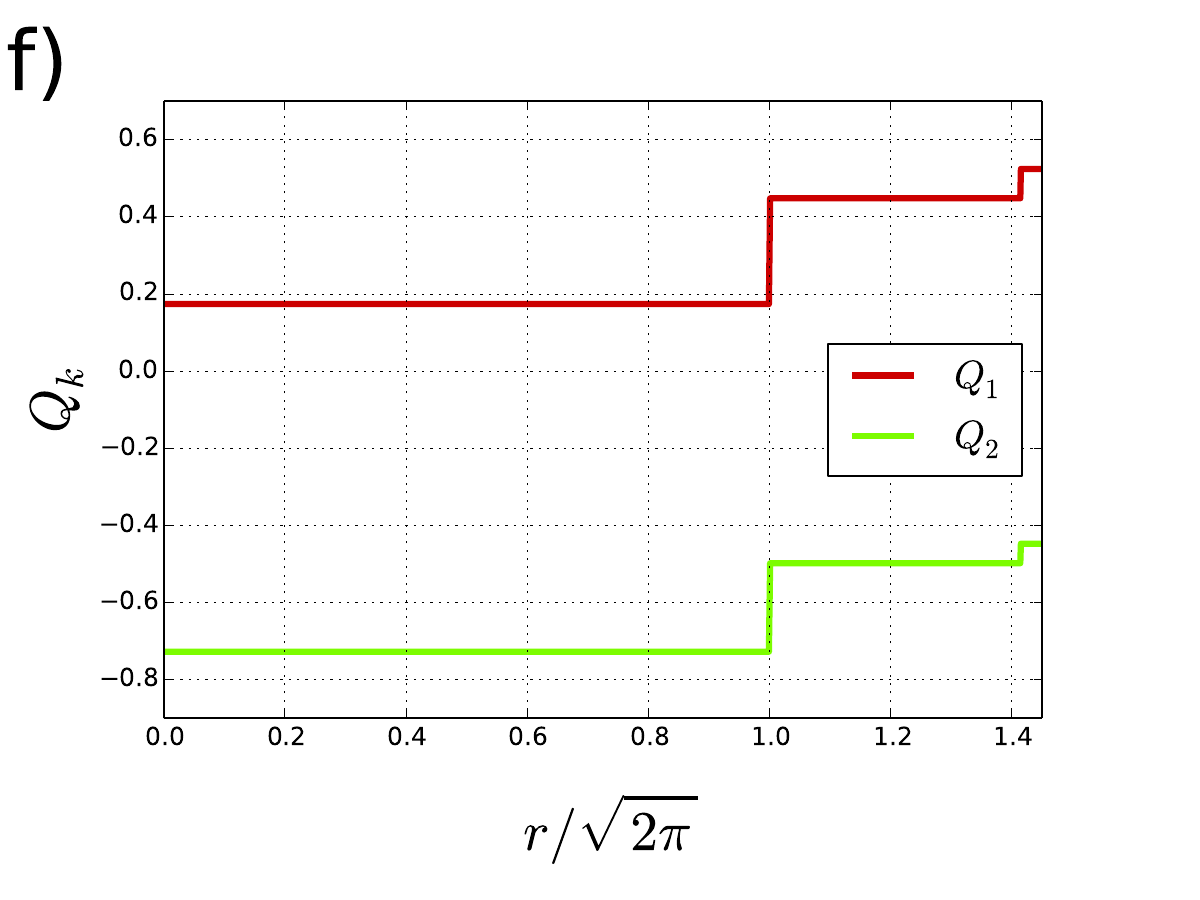}
	\includegraphics[width=0.245\textwidth]{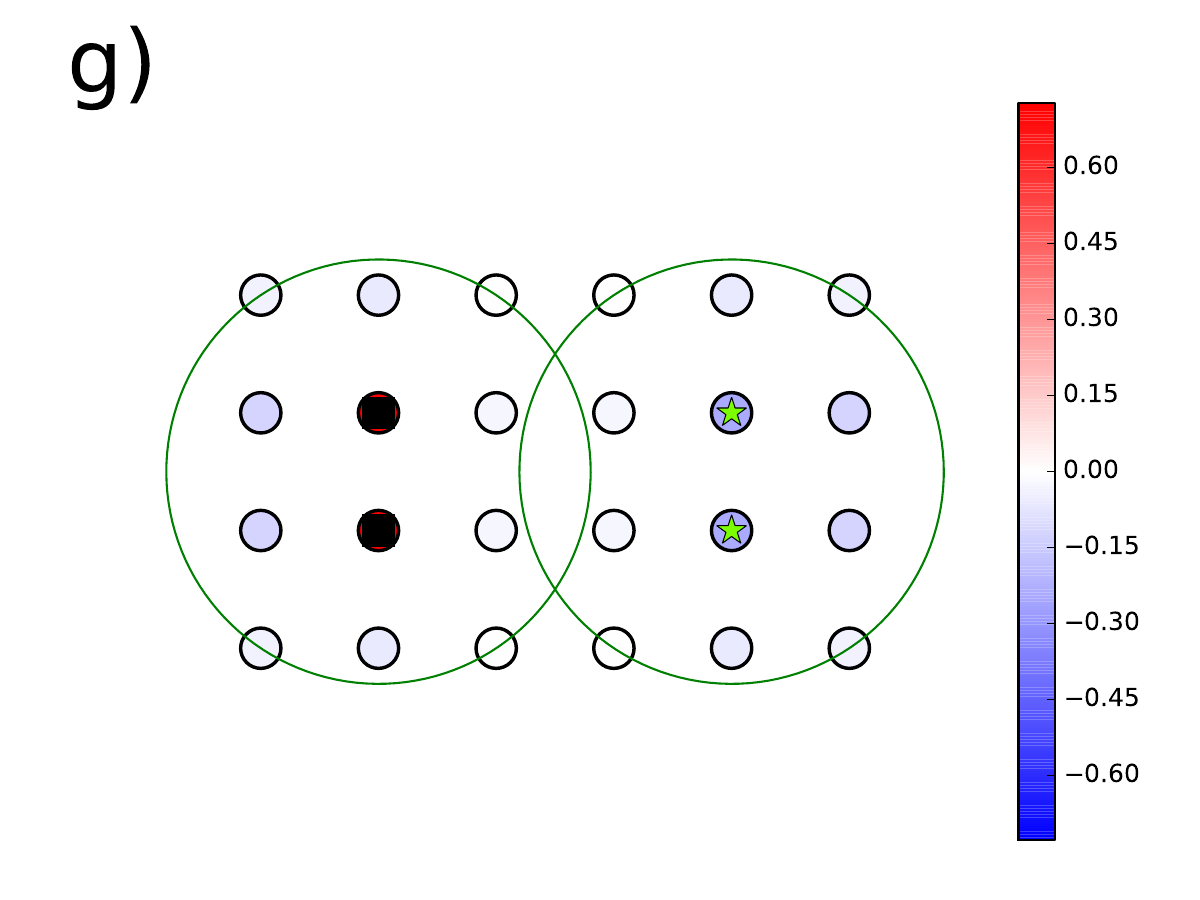}
	\includegraphics[width=0.245\textwidth]{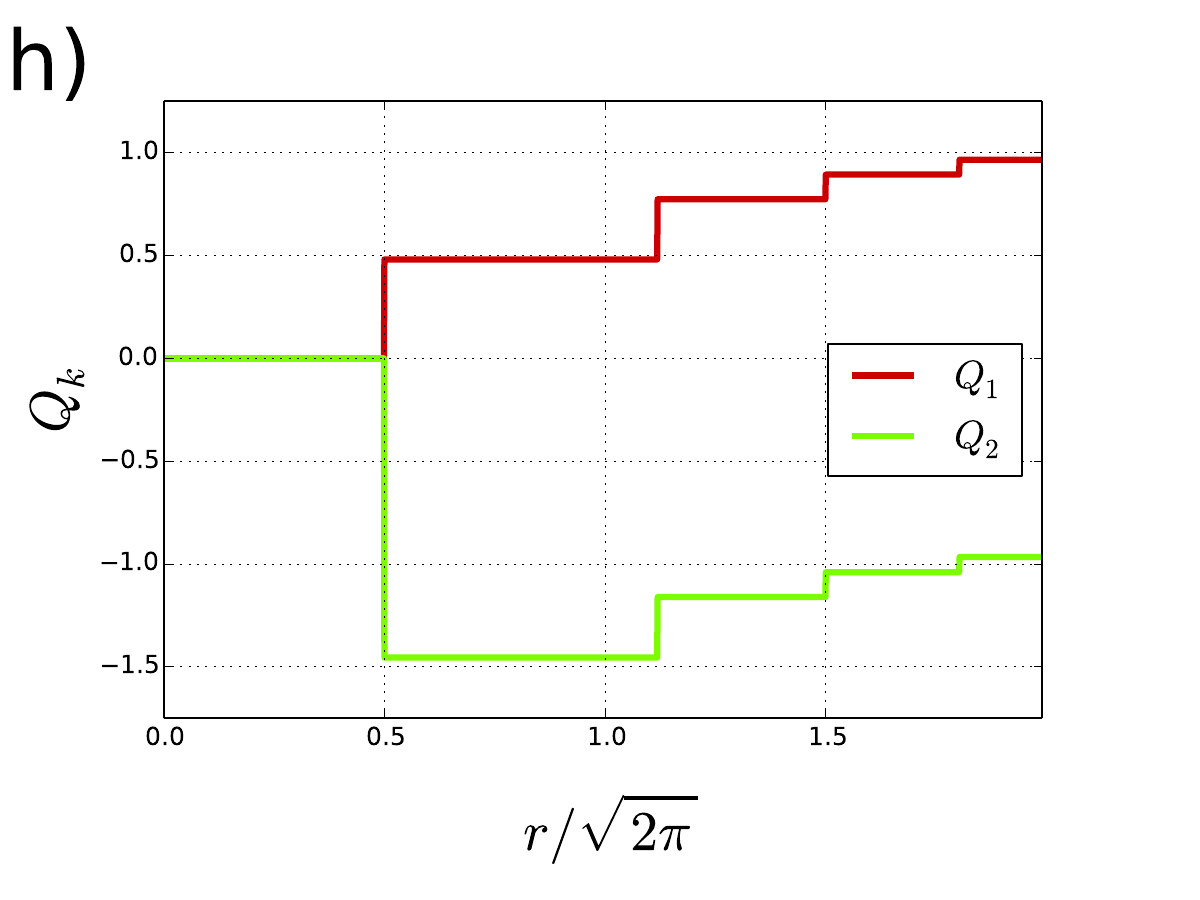}
	\caption{In $a), c), e)$ and $g)$ we symbolize lattice sites $(N = 24)$ by the circles.
		Quasiholes and quasielectrons are depicted by stars and squares, respectively.
		We place a quasihole potential $V$ and a quasielectron potential $-V$ on two different lattice sites in $a)$. Similarly we place two quasihole potentials (each with $V$) and two quasielectron potentials (each with $-V$) in $c)$.
		The density profile $\rho(z_i)$ 
		that is defined as the particle density difference
		between the states with and without anyons, is plotted with colorbar. We plot the excess charge distribution $Q_k$ as defined in \eqref{Excess_Charge}
		as a function of the radial distance from the anyons in $b)$ and $d)$ as marked by the circles respectively in $a)$ and $c)$. 
		In $e)$ and $g)$ we place one quasihole-one quasielectron (hence exploit \eqref{wavefunc_3} and \eqref{wavefunc_6}) and
		two quasiholes-two quasielectrons in the system (hence use \eqref{wavefunc_3} and \eqref{wavefunc_4}).
		The density profile $\rho(z_i)$ 
		is plotted with colorbar. We also plot the excess charge distribution $Q_k$ as a function of the radial distance in $f)$ and $h)$.
		We observe that in all cases the quasiparticles, both quasiholes and quasielectrons, are trapped with the charge of $\sim 0.5$ and $\sim -0.5$, respectively. Note that $b)$ and $f)$ are similar, although not exactly the same, and the same applies to $d)$ and $h)$
	}\label{Fig_KM_Ham_KM_MR}      
\end{figure*}

\section{Quasielectrons in the Kapit-Mueller model}\label{KM_Model}

In Ref.\onlinecite{Others47} Kapit and Mueller proposed a lattice model Hamiltonian, which realizes fractional quantum Hall physics at appropriate filling factors. This model exhibits an exact equivalence between a realistic lattice system and the lowest Landau level. The Hamiltonian is relatively simple, as it consists only of hopping terms and hardcore on-site interactions. Here we consider the filling factor, for which the ground state is known to be in the same topological phase as the bosonic Moore-Read state with $q = 1$. One may speculate if the shape of the quasielectrons computed from the analytical states is very specific to the analytical states or whether it is not. In the following, we therefore compare anyons in the Kapit-Mueller model to the anyons in the analytical states. 

The Hamiltonian is given by 
\begin{equation}\label{Ham_KM}
H_0 = - \sum_{j,k} J_{jk} e^{i \phi_{jk}} c^{\dagger}_j c_k + h.c.
\end{equation}
We take unit lattice spacing here. $c^{\dagger}_k (c_k)$ creates (annihilates) a hardcore boson at the $k$th lattice site at position $z_k$. 
We define $\xi_{jk} = z_j - z_k$ and $a = \text{Re}(\xi_{jk}), \ b = \text{Im}(\xi_{jk})$. 
The Peierls phase of the magnetic field is provided by $i \phi_{jk} = -\frac{\pi \phi}{2} \big( z_j \xi_{jk}^{*} - z^{*}_j \xi_{jk}  \big)$ with 
$\phi$ as the density of flux quanta per plaquette on a square lattice. 
We use Gaussian hopping couplings 
\begin{equation}
J_{jk} = G(\xi_{jk}) \text{exp} \big(  -\frac{\pi}{2} (1 - \phi) |\xi_{jk}|^2 \big)
\end{equation}
where $G(\xi_{jk}) = (-1)^{1+a+b+ab}$ to get robust fractional quantum Hall states as claimed in Ref.\onlinecite{Others47}.
 
We now incorporate a potential term $H_v$ in \eqref{Ham_KM} to localize \cite{Others46, Others53} quasiholes and quasielectrons.
This is done by providing an energy penalty to lattice sites to be occupied for the quasiholes and an energy penalty to lattice sites to be unoccupied for the quasielectrons. 
To trap anyons we define
\begin{equation}\label{Ham_KM_anyon}
\begin{split}
&H_v =  (n_a - n_b)V, \quad \text{Fig} \ \ref{Fig_KM_Ham_KM_MR}.a)\\&
H_v = (n_a + n_b - n_c - n_d)V, \quad \text{Fig} \ \ref{Fig_KM_Ham_KM_MR}.c)
\end{split}
\end{equation}
where $n_k = c^{\dagger}_k c_k$ is the number operator at the $k$th lattice site and $V$ is the strength of the potential at the $k$th lattice site to localize anyons.

We proceed to investigate density profile, charge and size of the anyons by doing an exact diagonalization study of the Hamiltonian $H_0 + H_v$ 
and by exploiting \eqref{Density_Profile} and \eqref{Excess_Charge}. 
We present our results on a small lattice system in Fig \ref{Fig_KM_Ham_KM_MR}.
In Fig \ref{Fig_KM_Ham_KM_MR}. $a)$ and $c)$ we show the density profile of a system of size $N = 24$ where we take a quasihole potential $(V)$ and a quasielectron potential $(-V)$ 
on one and two lattice sites respectively to localize anyons. We take the number of particles as $M = 4$. 
This together with a system size of $N = 24$ leads to $\phi = 1/6$ in both $a)$ and $e)$. Fig \ref{Fig_KM_Ham_KM_MR}. $b)$ and $f)$ display the excess charge 
distribution of the trapped anyons as a function of the radial distance from the anyons as marked by the circles respectively in $a)$ and $e)$. We find that one quasihole and one quasielectron are trapped in $a)$ and that two quasiholes and two quasielectrons are trapped in $c)$.
Each of the quasiholes and quasielectrons exhibits the charge of $\sim 0.5$ and $\sim -0.5$ respectively. 
Therefore the anyons are localized excitations and screened well. 

We will now investigate the analytical states on the same lattice for comparison. We display the results on a small lattice system in Fig \ref{Fig_KM_Ham_KM_MR} where $c)$ and $g)$ show the density profile in which we place one 
quasihole-one quasielectron and two quasiholes-two quasielectrons, respectively. 
We exploit Eq \eqref{particle_number} and take $M = 4, N = 24, q = 1$ (for bosons) and $\eta = 1/6$. Fig \ref{Fig_KM_Ham_KM_MR}. $d)$ and $h)$ 
display the excess charge distribution of the anyons for the region marked by the circles in $c)$ and $g)$, respectively. 
Each of the quasiholes and quasielectrons exhibits the charge of $ p_k/q \sim \pm 0.5$ 
concurrently. 
Therefore the anyons are localized in the systems. 
It is to be noted from Fig \ref{Fig_KM_Ham_KM_MR}. $b)$, $f)$ and Fig \ref{Fig_KM_Ham_KM_MR}. $d)$, $h)$ that the density profiles and the excess charge distributions 
are very similar. This shows a similarity between the properties of the ground state of the Kapit-Mueller model and of the analytical states.

\section{Conclusions} \label{concl}

Analytical models are very helpful to get insight in models with strong correlations in quantum many body systems and 
here we construct an anyonic model with analytical ground states and a parent Hamiltonian. 
We show that in lattice fractional quantum Hall models quasielectrons can be created in a similar way as quasiholes can be created. 
This approach results in simpler wave functions for quasielectrons both from analytical and numerical viewpoint than the continuous system. 
We have constructed Moore-Read states containing two quasielectrons, one quasihole-one quasielectron, four quasielectrons and two quasiholes-two quasielectrons. 
Detailed investigations of the density profile, charge, size and fractional braiding statistics show that the anyons are non-Abelian Ising anyons. 

We also investigated the Kapit-Mueller model on a small square lattice system and created quasielectrons in a similar fashion to the creation of quasiholes. We found that the quasielectrons in the Kapit-Mueller model have a shape similar to the shape of the quasielectrons in the analytical states. This shows that the analytical states are relevant also for models that are experimentally relevant.

In the future, it would be interesting to make similar studies for Fibonacci anyons in the Read-Rezayi states.\\

\section{Acknowledgments}
We thank Blazej Jaworowski. JW thanks NSF DMR 1306897 and NSF DMR 1056536 for partial support.

\bibliography{bibfile}

\begin{thebibliography}{27}%
\makeatletter
\providecommand \@ifxundefined [1]{%
 \@ifx{#1\undefined}
}%
\providecommand \@ifnum [1]{%
 \ifnum #1\expandafter \@firstoftwo
 \else \expandafter \@secondoftwo
 \fi
}%
\providecommand \@ifx [1]{%
 \ifx #1\expandafter \@firstoftwo
 \else \expandafter \@secondoftwo
 \fi
}%
\providecommand \natexlab [1]{#1}%
\providecommand \enquote  [1]{``#1''}%
\providecommand \bibnamefont  [1]{#1}%
\providecommand \bibfnamefont [1]{#1}%
\providecommand \citenamefont [1]{#1}%
\providecommand \href@noop [0]{\@secondoftwo}%
\providecommand \href [0]{\begingroup \@sanitize@url \@href}%
\providecommand \@href[1]{\@@startlink{#1}\@@href}%
\providecommand \@@href[1]{\endgroup#1\@@endlink}%
\providecommand \@sanitize@url [0]{\catcode `\\12\catcode `\$12\catcode
  `\&12\catcode `\#12\catcode `\^12\catcode `\_12\catcode `\%12\relax}%
\providecommand \@@startlink[1]{}%
\providecommand \@@endlink[0]{}%
\providecommand \url  [0]{\begingroup\@sanitize@url \@url }%
\providecommand \@url [1]{\endgroup\@href {#1}{\urlprefix }}%
\providecommand \urlprefix  [0]{URL }%
\providecommand \Eprint [0]{\href }%
\providecommand \doibase [0]{http://dx.doi.org/}%
\providecommand \selectlanguage [0]{\@gobble}%
\providecommand \bibinfo  [0]{\@secondoftwo}%
\providecommand \bibfield  [0]{\@secondoftwo}%
\providecommand \translation [1]{[#1]}%
\providecommand \BibitemOpen [0]{}%
\providecommand \bibitemStop [0]{}%
\providecommand \bibitemNoStop [0]{.\EOS\space}%
\providecommand \EOS [0]{\spacefactor3000\relax}%
\providecommand \BibitemShut  [1]{\csname bibitem#1\endcsname}%
\let\auto@bib@innerbib\@empty
\bibitem [{\citenamefont {Christian}\ and\ \citenamefont
  {Immanuel}(2017)}]{Others49}%
  \BibitemOpen
  \bibfield  {author} {\bibinfo {author} {\bibfnamefont {G.}~\bibnamefont
  {Christian}}\ and\ \bibinfo {author} {\bibfnamefont {B.}~\bibnamefont
  {Immanuel}},\ }\href {\doibase 10.1126/science.aal3837} {\bibfield  {journal}
  {\bibinfo  {journal} {Science}\ }\textbf {\bibinfo {volume} {357}},\ \bibinfo
  {pages} {995} (\bibinfo {year} {2017})}\BibitemShut {NoStop}%
\bibitem [{\citenamefont {Bergholtz}\ and\ \citenamefont
  {Liu}(2013)}]{Others51}%
  \BibitemOpen
  \bibfield  {author} {\bibinfo {author} {\bibfnamefont {E.~J.}\ \bibnamefont
  {Bergholtz}}\ and\ \bibinfo {author} {\bibfnamefont {Z.}~\bibnamefont
  {Liu}},\ }\href@noop {} {\bibfield  {journal} {\bibinfo  {journal} {Int. J.
  Mod. Phys. B}\ }\textbf {\bibinfo {volume} {27}},\ \bibinfo {pages} {1330017}
  (\bibinfo {year} {2013})}\BibitemShut {NoStop}%
\bibitem [{\citenamefont {Tang}\ \emph {et~al.}(2011)\citenamefont {Tang},
  \citenamefont {Mei},\ and\ \citenamefont {Wen}}]{Others48}%
  \BibitemOpen
  \bibfield  {author} {\bibinfo {author} {\bibfnamefont {E.}~\bibnamefont
  {Tang}}, \bibinfo {author} {\bibfnamefont {J.-W.}\ \bibnamefont {Mei}}, \
  and\ \bibinfo {author} {\bibfnamefont {X.-G.}\ \bibnamefont {Wen}},\ }\href
  {\doibase 10.1103/PhysRevLett.106.236802} {\bibfield  {journal} {\bibinfo
  {journal} {Phys. Rev. Lett.}\ }\textbf {\bibinfo {volume} {106}},\ \bibinfo
  {pages} {236802} (\bibinfo {year} {2011})}\BibitemShut {NoStop}%
\bibitem [{\citenamefont {Nayak}\ \emph {et~al.}(2008)\citenamefont {Nayak},
  \citenamefont {Simon}, \citenamefont {Stern}, \citenamefont {Freedman},\ and\
  \citenamefont {Das~Sarma}}]{C.Nayak9}%
  \BibitemOpen
  \bibfield  {author} {\bibinfo {author} {\bibfnamefont {C.}~\bibnamefont
  {Nayak}}, \bibinfo {author} {\bibfnamefont {S.~H.}\ \bibnamefont {Simon}},
  \bibinfo {author} {\bibfnamefont {A.}~\bibnamefont {Stern}}, \bibinfo
  {author} {\bibfnamefont {M.}~\bibnamefont {Freedman}}, \ and\ \bibinfo
  {author} {\bibfnamefont {S.}~\bibnamefont {Das~Sarma}},\ }\href {\doibase
  10.1103/RevModPhys.80.1083} {\bibfield  {journal} {\bibinfo  {journal} {Rev.
  Mod. Phys.}\ }\textbf {\bibinfo {volume} {80}},\ \bibinfo {pages} {1083}
  (\bibinfo {year} {2008})}\BibitemShut {NoStop}%
\bibitem [{\citenamefont {Bonderson}\ \emph {et~al.}(2011)\citenamefont
  {Bonderson}, \citenamefont {Gurarie},\ and\ \citenamefont
  {Nayak}}]{C.Nayak8}%
  \BibitemOpen
  \bibfield  {author} {\bibinfo {author} {\bibfnamefont {P.}~\bibnamefont
  {Bonderson}}, \bibinfo {author} {\bibfnamefont {V.}~\bibnamefont {Gurarie}},
  \ and\ \bibinfo {author} {\bibfnamefont {C.}~\bibnamefont {Nayak}},\ }\href
  {\doibase 10.1103/PhysRevB.83.075303} {\bibfield  {journal} {\bibinfo
  {journal} {Phys. Rev. B}\ }\textbf {\bibinfo {volume} {83}},\ \bibinfo
  {pages} {075303} (\bibinfo {year} {2011})}\BibitemShut {NoStop}%
\bibitem [{\citenamefont {Wu}\ \emph {et~al.}(2014)\citenamefont {Wu},
  \citenamefont {Estienne}, \citenamefont {Regnault},\ and\ \citenamefont
  {Bernevig}}]{BAB14}%
  \BibitemOpen
  \bibfield  {author} {\bibinfo {author} {\bibfnamefont {Y.-L.}\ \bibnamefont
  {Wu}}, \bibinfo {author} {\bibfnamefont {B.}~\bibnamefont {Estienne}},
  \bibinfo {author} {\bibfnamefont {N.}~\bibnamefont {Regnault}}, \ and\
  \bibinfo {author} {\bibfnamefont {B.~A.}\ \bibnamefont {Bernevig}},\ }\href
  {\doibase 10.1103/PhysRevLett.113.116801} {\bibfield  {journal} {\bibinfo
  {journal} {Phys. Rev. Lett.}\ }\textbf {\bibinfo {volume} {113}},\ \bibinfo
  {pages} {116801} (\bibinfo {year} {2014})}\BibitemShut {NoStop}%
\bibitem [{\citenamefont {Arovas}\ \emph {et~al.}(1984)\citenamefont {Arovas},
  \citenamefont {Schrieffer},\ and\ \citenamefont {Wilczek}}]{DA1}%
  \BibitemOpen
  \bibfield  {author} {\bibinfo {author} {\bibfnamefont {D.}~\bibnamefont
  {Arovas}}, \bibinfo {author} {\bibfnamefont {J.~R.}\ \bibnamefont
  {Schrieffer}}, \ and\ \bibinfo {author} {\bibfnamefont {F.}~\bibnamefont
  {Wilczek}},\ }\href {\doibase 10.1103/PhysRevLett.53.722} {\bibfield
  {journal} {\bibinfo  {journal} {Phys. Rev. Lett.}\ }\textbf {\bibinfo
  {volume} {53}},\ \bibinfo {pages} {722} (\bibinfo {year} {1984})}\BibitemShut
  {NoStop}%
\bibitem [{\citenamefont {Jaworowski}\ \emph {et~al.}(2018)\citenamefont
  {Jaworowski}, \citenamefont {Regnault},\ and\ \citenamefont
  {Liu}}]{Others52}%
  \BibitemOpen
  \bibfield  {author} {\bibinfo {author} {\bibfnamefont {B.}~\bibnamefont
  {Jaworowski}}, \bibinfo {author} {\bibfnamefont {N.}~\bibnamefont
  {Regnault}}, \ and\ \bibinfo {author} {\bibfnamefont {Z.}~\bibnamefont
  {Liu}},\ }\href@noop {} {\bibfield  {journal} {\bibinfo  {journal}
  {arXiv:1810.03458}\ } (\bibinfo {year} {2018})}\BibitemShut {NoStop}%
\bibitem [{\citenamefont {Hansson}\ \emph
  {et~al.}(2009{\natexlab{a}})\citenamefont {Hansson}, \citenamefont
  {Hermanns}, \citenamefont {Regnault},\ and\ \citenamefont {Viefers}}]{NR3}%
  \BibitemOpen
  \bibfield  {author} {\bibinfo {author} {\bibfnamefont {T.~H.}\ \bibnamefont
  {Hansson}}, \bibinfo {author} {\bibfnamefont {M.}~\bibnamefont {Hermanns}},
  \bibinfo {author} {\bibfnamefont {N.}~\bibnamefont {Regnault}}, \ and\
  \bibinfo {author} {\bibfnamefont {S.}~\bibnamefont {Viefers}},\ }\href
  {\doibase 10.1103/PhysRevLett.102.166805} {\bibfield  {journal} {\bibinfo
  {journal} {Phys. Rev. Lett.}\ }\textbf {\bibinfo {volume} {102}},\ \bibinfo
  {pages} {166805} (\bibinfo {year} {2009}{\natexlab{a}})}\BibitemShut
  {NoStop}%
\bibitem [{\citenamefont {Rodriguez}\ \emph {et~al.}(2012)\citenamefont
  {Rodriguez}, \citenamefont {Sterdyniak}, \citenamefont {Hermanns},
  \citenamefont {Slingerland},\ and\ \citenamefont {Regnault}}]{NR5}%
  \BibitemOpen
  \bibfield  {author} {\bibinfo {author} {\bibfnamefont {I.~D.}\ \bibnamefont
  {Rodriguez}}, \bibinfo {author} {\bibfnamefont {A.}~\bibnamefont
  {Sterdyniak}}, \bibinfo {author} {\bibfnamefont {M.}~\bibnamefont
  {Hermanns}}, \bibinfo {author} {\bibfnamefont {J.~K.}\ \bibnamefont
  {Slingerland}}, \ and\ \bibinfo {author} {\bibfnamefont {N.}~\bibnamefont
  {Regnault}},\ }\href {\doibase 10.1103/PhysRevB.85.035128} {\bibfield
  {journal} {\bibinfo  {journal} {Phys. Rev. B}\ }\textbf {\bibinfo {volume}
  {85}},\ \bibinfo {pages} {035128} (\bibinfo {year} {2012})}\BibitemShut
  {NoStop}%
\bibitem [{\citenamefont {Hansson}\ \emph
  {et~al.}(2009{\natexlab{b}})\citenamefont {Hansson}, \citenamefont
  {Hermanns},\ and\ \citenamefont {Viefers}}]{Others41}%
  \BibitemOpen
  \bibfield  {author} {\bibinfo {author} {\bibfnamefont {T.~H.}\ \bibnamefont
  {Hansson}}, \bibinfo {author} {\bibfnamefont {M.}~\bibnamefont {Hermanns}}, \
  and\ \bibinfo {author} {\bibfnamefont {S.}~\bibnamefont {Viefers}},\ }\href
  {\doibase 10.1103/PhysRevB.80.165330} {\bibfield  {journal} {\bibinfo
  {journal} {Phys. Rev. B}\ }\textbf {\bibinfo {volume} {80}},\ \bibinfo
  {pages} {165330} (\bibinfo {year} {2009}{\natexlab{b}})}\BibitemShut
  {NoStop}%
\bibitem [{\citenamefont {Bernevig}\ and\ \citenamefont
  {Haldane}(2009)}]{BAB5}%
  \BibitemOpen
  \bibfield  {author} {\bibinfo {author} {\bibfnamefont {B.~A.}\ \bibnamefont
  {Bernevig}}\ and\ \bibinfo {author} {\bibfnamefont {F.~D.~M.}\ \bibnamefont
  {Haldane}},\ }\href {\doibase 10.1103/PhysRevLett.102.066802} {\bibfield
  {journal} {\bibinfo  {journal} {Phys. Rev. Lett.}\ }\textbf {\bibinfo
  {volume} {102}},\ \bibinfo {pages} {066802} (\bibinfo {year}
  {2009})}\BibitemShut {NoStop}%
\bibitem [{\citenamefont {Yang}\ and\ \citenamefont {Haldane}(2014)}]{FDMH1}%
  \BibitemOpen
  \bibfield  {author} {\bibinfo {author} {\bibfnamefont {B.}~\bibnamefont
  {Yang}}\ and\ \bibinfo {author} {\bibfnamefont {F.~D.~M.}\ \bibnamefont
  {Haldane}},\ }\href {\doibase 10.1103/PhysRevLett.112.026804} {\bibfield
  {journal} {\bibinfo  {journal} {Phys. Rev. Lett.}\ }\textbf {\bibinfo
  {volume} {112}},\ \bibinfo {pages} {026804} (\bibinfo {year}
  {2014})}\BibitemShut {NoStop}%
\bibitem [{\citenamefont {Kj{\"a}ll}\ \emph {et~al.}(2018)\citenamefont
  {Kj{\"a}ll}, \citenamefont {Ardonne}, \citenamefont {Dwivedi}, \citenamefont
  {Hermanns},\ and\ \citenamefont {Hansson}}]{Others42}%
  \BibitemOpen
  \bibfield  {author} {\bibinfo {author} {\bibfnamefont {J.}~\bibnamefont
  {Kj{\"a}ll}}, \bibinfo {author} {\bibfnamefont {E.}~\bibnamefont {Ardonne}},
  \bibinfo {author} {\bibfnamefont {V.}~\bibnamefont {Dwivedi}}, \bibinfo
  {author} {\bibfnamefont {M.}~\bibnamefont {Hermanns}}, \ and\ \bibinfo
  {author} {\bibfnamefont {T.~H.}\ \bibnamefont {Hansson}},\ }\href
  {http://stacks.iop.org/1742-5468/2018/i=5/a=053101} {\bibfield  {journal}
  {\bibinfo  {journal} {Journal of Statistical Mechanics: Theory and
  Experiment}\ }\textbf {\bibinfo {volume} {2018}},\ \bibinfo {pages} {053101}
  (\bibinfo {year} {2018})}\BibitemShut {NoStop}%
\bibitem [{\citenamefont {Greiter}\ \emph {et~al.}(2016)\citenamefont
  {Greiter}, \citenamefont {Schnells},\ and\ \citenamefont
  {Thomale}}]{Others43}%
  \BibitemOpen
  \bibfield  {author} {\bibinfo {author} {\bibfnamefont {M.}~\bibnamefont
  {Greiter}}, \bibinfo {author} {\bibfnamefont {V.}~\bibnamefont {Schnells}}, \
  and\ \bibinfo {author} {\bibfnamefont {R.}~\bibnamefont {Thomale}},\ }\href
  {\doibase 10.1103/PhysRevB.93.245156} {\bibfield  {journal} {\bibinfo
  {journal} {Phys. Rev. B}\ }\textbf {\bibinfo {volume} {93}},\ \bibinfo
  {pages} {245156} (\bibinfo {year} {2016})}\BibitemShut {NoStop}%
\bibitem [{\citenamefont {Jeon}\ and\ \citenamefont {Jain}(2003)}]{JKJ7}%
  \BibitemOpen
  \bibfield  {author} {\bibinfo {author} {\bibfnamefont {G.~S.}\ \bibnamefont
  {Jeon}}\ and\ \bibinfo {author} {\bibfnamefont {J.~K.}\ \bibnamefont
  {Jain}},\ }\href {\doibase 10.1103/PhysRevB.68.165346} {\bibfield  {journal}
  {\bibinfo  {journal} {Phys. Rev. B}\ }\textbf {\bibinfo {volume} {68}},\
  \bibinfo {pages} {165346} (\bibinfo {year} {2003})}\BibitemShut {NoStop}%
\bibitem [{\citenamefont {Nielsen}\ \emph {et~al.}(2018)\citenamefont
  {Nielsen}, \citenamefont {Glasser},\ and\ \citenamefont {Rodriguez}}]{Anne2}%
  \BibitemOpen
  \bibfield  {author} {\bibinfo {author} {\bibfnamefont {A.~E.~B.}\
  \bibnamefont {Nielsen}}, \bibinfo {author} {\bibfnamefont {I.}~\bibnamefont
  {Glasser}}, \ and\ \bibinfo {author} {\bibfnamefont {I.~D.}\ \bibnamefont
  {Rodriguez}},\ }\href@noop {} {\bibfield  {journal} {\bibinfo  {journal} {New
  Journal of Physics}\ }\textbf {\bibinfo {volume} {20}},\ \bibinfo {pages}
  {033029} (\bibinfo {year} {2018})}\BibitemShut {NoStop}%
\bibitem [{\citenamefont {Moore}\ and\ \citenamefont
  {Read}(1991)}]{Moore-Read1}%
  \BibitemOpen
  \bibfield  {author} {\bibinfo {author} {\bibfnamefont {G.}~\bibnamefont
  {Moore}}\ and\ \bibinfo {author} {\bibfnamefont {N.}~\bibnamefont {Read}},\
  }\href {\doibase https://doi.org/10.1016/0550-3213(91)90407-O} {\bibfield
  {journal} {\bibinfo  {journal} {Nuclear Physics B}\ }\textbf {\bibinfo
  {volume} {360}},\ \bibinfo {pages} {362 } (\bibinfo {year}
  {1991})}\BibitemShut {NoStop}%
\bibitem [{\citenamefont {Kapit}\ and\ \citenamefont
  {Mueller}(2010)}]{Others47}%
  \BibitemOpen
  \bibfield  {author} {\bibinfo {author} {\bibfnamefont {E.}~\bibnamefont
  {Kapit}}\ and\ \bibinfo {author} {\bibfnamefont {E.}~\bibnamefont
  {Mueller}},\ }\href {\doibase 10.1103/PhysRevLett.105.215303} {\bibfield
  {journal} {\bibinfo  {journal} {Phys. Rev. Lett.}\ }\textbf {\bibinfo
  {volume} {105}},\ \bibinfo {pages} {215303} (\bibinfo {year}
  {2010})}\BibitemShut {NoStop}%
\bibitem [{\citenamefont {Manna}\ \emph {et~al.}(2018)\citenamefont {Manna},
  \citenamefont {Wildeboer}, \citenamefont {Sierra},\ and\ \citenamefont
  {Nielsen}}]{Anne8}%
  \BibitemOpen
  \bibfield  {author} {\bibinfo {author} {\bibfnamefont {S.}~\bibnamefont
  {Manna}}, \bibinfo {author} {\bibfnamefont {J.}~\bibnamefont {Wildeboer}},
  \bibinfo {author} {\bibfnamefont {G.}~\bibnamefont {Sierra}}, \ and\ \bibinfo
  {author} {\bibfnamefont {A.~E.~B.}\ \bibnamefont {Nielsen}},\ }\href
  {\doibase 10.1103/PhysRevB.98.165147} {\bibfield  {journal} {\bibinfo
  {journal} {Phys. Rev. B}\ }\textbf {\bibinfo {volume} {98}},\ \bibinfo
  {pages} {165147} (\bibinfo {year} {2018})}\BibitemShut {NoStop}%
\bibitem [{\citenamefont {Ardonne}\ and\ \citenamefont {Sierra}()}]{German1}%
  \BibitemOpen
  \bibfield  {author} {\bibinfo {author} {\bibfnamefont {E.}~\bibnamefont
  {Ardonne}}\ and\ \bibinfo {author} {\bibfnamefont {G.}~\bibnamefont
  {Sierra}},\ }\href {http://stacks.iop.org/1751-8121/43/i=50/a=505402}
  {\bibfield  {journal} {\bibinfo  {journal} {Journal of Physics A:
  Mathematical and Theoretical}\ }\textbf {\bibinfo {volume} {43}},\ \bibinfo
  {pages} {505402}}\BibitemShut {NoStop}%
\bibitem [{\citenamefont {Willett}\ \emph {et~al.}(2009)\citenamefont
  {Willett}, \citenamefont {Pfeiffer},\ and\ \citenamefont {West}}]{Others1}%
  \BibitemOpen
  \bibfield  {author} {\bibinfo {author} {\bibfnamefont {R.~L.}\ \bibnamefont
  {Willett}}, \bibinfo {author} {\bibfnamefont {L.~N.}\ \bibnamefont
  {Pfeiffer}}, \ and\ \bibinfo {author} {\bibfnamefont {K.~W.}\ \bibnamefont
  {West}},\ }\href {\doibase 10.1073/pnas.0812599106} {\bibfield  {journal}
  {\bibinfo  {journal} {PNAS}\ }\textbf {\bibinfo {volume} {106}} (\bibinfo
  {year} {2009}),\ 10.1073/pnas.0812599106}\BibitemShut {NoStop}%
\bibitem [{\citenamefont {Vivek~Venkatachalam}\ and\ \citenamefont
  {West}(2011)}]{Others5}%
  \BibitemOpen
  \bibfield  {author} {\bibinfo {author} {\bibfnamefont {L.~P.}\ \bibnamefont
  {Vivek~Venkatachalam}, \bibfnamefont {Amir~Yacoby}}\ and\ \bibinfo {author}
  {\bibfnamefont {K.}~\bibnamefont {West}},\ }\href {\doibase
  10.1038/nature09680} {\bibfield  {journal} {\bibinfo  {journal} {Nature}\
  }\textbf {\bibinfo {volume} {469}},\ \bibinfo {pages} {185} (\bibinfo {year}
  {2011})}\BibitemShut {NoStop}%
\bibitem [{\citenamefont {Glasser}\ \emph {et~al.}(2015)\citenamefont
  {Glasser}, \citenamefont {Cirac}, \citenamefont {Sierra},\ and\ \citenamefont
  {Nielsen}}]{Anne4}%
  \BibitemOpen
  \bibfield  {author} {\bibinfo {author} {\bibfnamefont {I.}~\bibnamefont
  {Glasser}}, \bibinfo {author} {\bibfnamefont {J.~I.}\ \bibnamefont {Cirac}},
  \bibinfo {author} {\bibfnamefont {G.}~\bibnamefont {Sierra}}, \ and\ \bibinfo
  {author} {\bibfnamefont {A.~E.~B.}\ \bibnamefont {Nielsen}},\ }\href
  {http://stacks.iop.org/1367-2630/17/i=8/a=082001} {\bibfield  {journal}
  {\bibinfo  {journal} {New Journal of Physics}\ }\textbf {\bibinfo {volume}
  {17}},\ \bibinfo {pages} {082001} (\bibinfo {year} {2015})}\BibitemShut
  {NoStop}%
\bibitem [{\citenamefont {Chen}\ \emph {et~al.}(2018)\citenamefont {Chen},
  \citenamefont {Vanderstraeten}, \citenamefont {Capponi},\ and\ \citenamefont
  {Poilblanc}}]{Others50}%
  \BibitemOpen
  \bibfield  {author} {\bibinfo {author} {\bibfnamefont {J.-Y.}\ \bibnamefont
  {Chen}}, \bibinfo {author} {\bibfnamefont {L.}~\bibnamefont
  {Vanderstraeten}}, \bibinfo {author} {\bibfnamefont {S.}~\bibnamefont
  {Capponi}}, \ and\ \bibinfo {author} {\bibfnamefont {D.}~\bibnamefont
  {Poilblanc}},\ }\href@noop {} {\bibfield  {journal} {\bibinfo  {journal}
  {arXiv:1807.04385}\ } (\bibinfo {year} {2018})}\BibitemShut {NoStop}%
\bibitem [{\citenamefont {Kapit}\ \emph {et~al.}(2012)\citenamefont {Kapit},
  \citenamefont {Ginsparg},\ and\ \citenamefont {Mueller}}]{Others46}%
  \BibitemOpen
  \bibfield  {author} {\bibinfo {author} {\bibfnamefont {E.}~\bibnamefont
  {Kapit}}, \bibinfo {author} {\bibfnamefont {P.}~\bibnamefont {Ginsparg}}, \
  and\ \bibinfo {author} {\bibfnamefont {E.}~\bibnamefont {Mueller}},\ }\href
  {\doibase 10.1103/PhysRevLett.108.066802} {\bibfield  {journal} {\bibinfo
  {journal} {Phys. Rev. Lett.}\ }\textbf {\bibinfo {volume} {108}},\ \bibinfo
  {pages} {066802} (\bibinfo {year} {2012})}\BibitemShut {NoStop}%
\bibitem [{\citenamefont {Raciunas}\ \emph {et~al.}(2018)\citenamefont
  {Raciunas}, \citenamefont {Unal}, \citenamefont {Anisimovas},\ and\
  \citenamefont {Eckardt}}]{Others53}%
  \BibitemOpen
  \bibfield  {author} {\bibinfo {author} {\bibfnamefont {M.}~\bibnamefont
  {Raciunas}}, \bibinfo {author} {\bibfnamefont {F.~N.}\ \bibnamefont {Unal}},
  \bibinfo {author} {\bibfnamefont {E.}~\bibnamefont {Anisimovas}}, \ and\
  \bibinfo {author} {\bibfnamefont {A.}~\bibnamefont {Eckardt}},\ }\href@noop
  {} {\bibfield  {journal} {\bibinfo  {journal} {arXiv:1804.02002}\ } (\bibinfo
  {year} {2018})}\BibitemShut {NoStop}%
\end{thebibliography}%
\end{document}